\RequirePackage{fix-cm}
\documentclass[twocolumn]{svjour3}          
\smartqed  
\usepackage[utf8]{inputenc}
\usepackage[abbreviations]{foreign}  
\usepackage{tabularx, makecell, booktabs}
\usepackage{tikz}
\usepackage[hidelinks]{hyperref}
\usepackage{comment}
\usepackage{subfigure}
\usepackage{xifthen}
\usepackage{hyperref}
\usepackage{xurl}
\usepackage{pifont}
\usepackage[justification=centering]{caption}
\usepackage{enumitem}
\usepackage{geometry}
\usepackage{rotating}
\usepackage{multirow}
\usepackage{xcolor}
\usepackage{orcidlink}
\usepackage{amssymb}

\begin{document}
\emergencystretch 3em

\title{TEE-based Key-Value Stores : a Survey}

\author{Aghiles Ait Messaoud\orcidlink{0000-0003-4657-5179} \and 
        Sonia Ben Mokhtar\orcidlink{0000-0003-2821-7714} \and
        Anthony Simonet-Boulogne\orcidlink{0000-0002-4072-8886}
}

\institute{A. Ait Messaoud \at
        LIRIS, iExec Blockchain Tech, Lyon, France \\
        \email{aghiles.ait-messaoud@insa-lyon.fr}     
    \and
       S. Ben Mokhtar \at
        LIRIS, CNRS, Lyon, France \\
        \email{sonia.benmokhtar@insa-lyon.fr}
    \and
        A. Simonet-Boulogne \at
        iExec Blockchain Tech, Lyon, France\\
        \email{anthony.simonet-boulogne@iex.ec}
}

\newif\ifshowcomments
\showcommentstrue

\ifshowcomments
  \newcommand{\mynote}[3]{%
    \fbox{\textbf{\sffamily\scriptsize #1}}%
    {\small$\blacktriangleright$\textsf{\emph{\color{#3}{#2}}}$\blacktriangleleft$}%
  }
\else
  \excludecomment{mynote}
\fi

\newcommand{\tabitem}{~~\llap{\textbullet}~~}
\newcommand{\sonia}[1]{\mynote{Sonia}{#1}{magenta}}
\newcommand{\anthony}[1]{\mynote{Anthony}{#1}{orange}}
\newcommand{\aghiles}[1]{\mynote{Aghiles}{#1}{blue}}

\newcommand*\circled[1]{\tikz[baseline=(char.base)]{
            \node[shape=circle,fill,inner sep=1pt] (char) {\textcolor{white}{#1}};}}

\newcommand{\cmark}{\ding{51}}
\newcommand{\xmark}{\ding{55}}

\newcommand*\colourcheck[1]{%
  \expandafter\newcommand\csname #1check\endcsname{\textcolor{#1}{\ding{51}}}%
}
\newcommand*\colourcross[1]{%
  \expandafter\newcommand\csname #1cross\endcsname{\textcolor{#1}{\ding{55}}}%
}

\colourcheck{green}
\colourcheck{gray}
\colourcross{red}

\maketitle

\begin{abstract}
Key-Value Stores (KVSs) are No-SQL databases that store data as key-value pairs and have gained popularity due to their simplicity, scalability, and fast retrieval capabilities. However, storing sensitive data in KVSs requires strong security properties to prevent data leakage and unauthorized tampering. While software (SW)-based encryption techniques are commonly used to maintain data confidentiality and integrity, they suffer from several drawbacks. They strongly assume trust in the hosting system stack and do not secure data during processing unless using performance-heavy techniques (\eg homomorphic encryption). Alternatively, Trusted Execution Environments (TEEs) provide a solution that enforces the confidentiality and integrity of code and data at the CPU level, allowing users to build trusted applications in an untrusted environment. They also secure data in use by providing an encapsulated processing environment called \textit{enclave}. Nevertheless, TEEs come with their own set of drawbacks, including performance issues due to memory size limitations and CPU context switching. This paper examines the state of the art in TEE-based confidential KVSs and highlights common design strategies used in KVSs to leverage TEE security features while overcoming their inherent limitations.
This work aims to provide a comprehensive understanding of the use of TEEs in KVSs and to identify research directions for future work.
\keywords{Key-Value-Store \and TEE \and SGX \and Privacy}
\end{abstract}

\section{Introduction}
Key-value stores (KVSs) like Redis~\cite{redis}, DynamoDB~\cite{dynamo_db}, and RocksDB~\cite{web:rocksdb} are NoSQL databases that have gained popularity since the cloud shift post-2010. KVSs provide a simple, fast, and scalable storage solution for unstructured or semi-structured data by mapping each key to a unique value. They store various types of data, including sensitive and confidential information such as session details~\cite{dynamo_db}, private cryptographic keys~\cite{aws-kms}, and wallet data~\cite{wallet}, necessitating strong confidentiality and integrity protections.

Traditional solutions for securing KVSs include software-based encrypted databases like~\cite{enckv,bigsecret,cryptDB,monomi}, which keep data encrypted along with verification metadata. Clear-text data is accessible only to those with the decryption key, and data integrity is ensured through digests or MACs that are compared to known references. However, these methods merely shift the problem to securing cryptographic keys. If the cryptographic keys are kept by the data owner, only the latter can use the outsourced data, complicating cloud-based data sharing unless keys are shared with authorized clients, risking leakage. If the cryptographic keys are managed by the database server or a Key Management Service (KMS), trust must be placed in their system administrators and software stack to not misuse them. Building secure KVSs for confidential data under such conditions implies strong trust assumptions. Moreover, software-based encryption techniques protect data at rest and in transit but require expensive methods like homomorphic encryption~\cite{homorphic-encryption} or verifiable computation~\cite{vc-1,vc-2,vc-3,vc-4} to protect data during processing.

An alternative to software-based encryption is using Trusted Execution Environments (TEEs)\cite{tee}. TEEs provide confidentiality and integrity at the hardware level by creating isolated secure environments, or enclaves, ensuring more robust isolation than virtualization-based techniques\cite{Jithin2014VirtualMI,subvirt}. Therefore, TEEs shift the trust in computation from the server software stack to the enclave code and its underlying CPU chip. TEEs are utilized, among other things, in secure mobile platforms~\cite{samsung_knox}, secure payment systems~\cite{web:samsung_pay}, or even privacy-preserving federated learning~\cite{bonawitz2019towards,ppfl,gradsec,shuffle_fl}, becoming central to securing sensitive data across various fields.

Despite several TEE implementations~\cite{cryptoeprint:sgx,whitepaper:kaplan2016amd,pinto2019demystifying}, SGX~\cite{cryptoeprint:sgx} is the most widely deployed due to Intel's market share and investment in SGX development and tools. For instance, compared to ARM TrustZone~\cite{pinto2019demystifying}, SGX offers built-in support for remote attestation~\cite{web:sgx-remote-attestation}, allowing verification of enclave authenticity and code integrity. However, SGX suffers from performance issues due to CPU context switching and its limited secure memory, which leads to in-enclave page faults and, thus, costly page swapping. To balance security and performance, SGX-based KVSs adopt novel designs and optimization techniques.

In this study, we examine TEE-based No-SQL KVSs and how they leverage the benefits of TEEs while addressing their limitations, especially those of SGX. All the existing TEE-based KVSs we could survey leverage Intel SGX. We believe this is due to the maturity and specific features of the Intel implementation (\eg resilience against physical attacks). We exclude general-purpose frameworks such as Scone~\cite{scone}, Gramine~\cite{graphene-sgx}, Haven~\cite{haven}, Panoply~\cite{panoply}, and SGX-LKL~\cite{sgx-lkl}, which port existing codebases to enclaves with minimal effort and no optimization. Instead, we focus on KVSs with novel designs that address SGX limitations. We also target NoSQL KVSs and exclude SQL databases such as~\cite{veridb,enclavedb,trusteddb,opaque}, as they serve different purposes, such as handling relational data (\eg for data analytics), and do not prioritize performance (including scalability). Moreover, these systems require a deeper literature review to address their specificities (parsers, optimizers, \etc). 

\begin{figure*}
    \centering
    \includegraphics[width=0.95\textwidth 
    ]{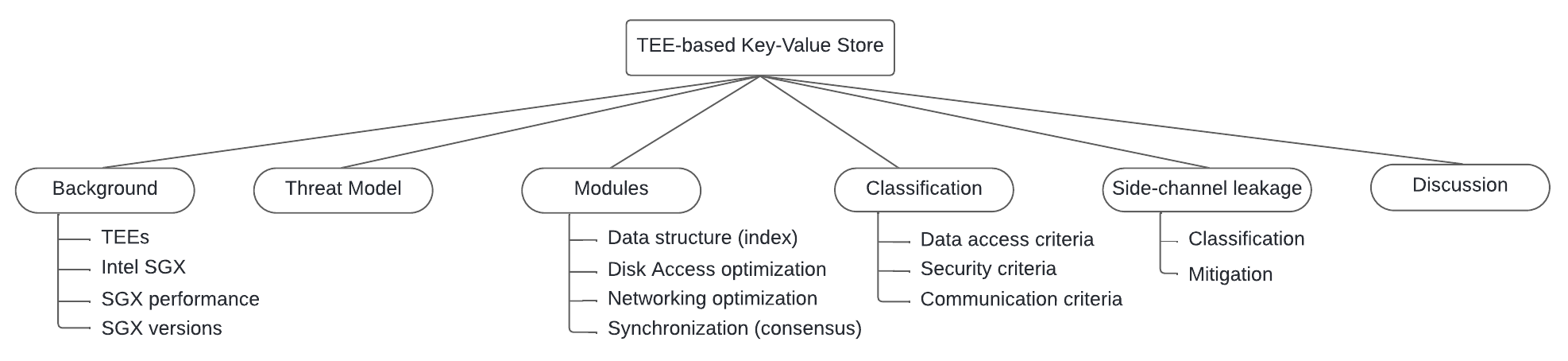}
    \caption{Survey flow}
    \label{fig:paper_flow}
\end{figure*}

The flow of this survey is depicted in Figure~\ref{fig:paper_flow}.
Section~\ref{sec:background} delves into the TEE concept, SGX implementation, limitations, and versions.
Section~\ref{sec:threat-model} formalizes the adopted threat model for TEE-based KVSs.
Section~\ref{sec:kvs-modules} synthesizes a modular architecture for the surveyed SGX-based KVSs and explains the roles and instantiations of each module.
Section~\ref{sec:classification} classifies SGX-based KVSs based on relevant criteria and implementation strategies.
Section~\ref{sec:sca} overviews the problem of side-channel attacks (SCAs) in SGX applications.
Section~\ref{sec:discussion} discusses open challenges related to SGX.
Finally, Section~\ref{sec:conclusion} concludes the work. 

\section{Background}\label{sec:background}
    \subsection{Trusted Execution Environments}
       In this section, we define TEEs and provide a comparison between major TEE implementations with respect to their security features.
        \subsubsection{Definition}
       TEEs are specialized environments implemented on commodity processors to protect applications (or portions of applications) in a secure environment isolated from other processes and the — potentially malicious — operating system (OS). They are turnkey solutions for securing applications, with early implementations appearing in the mid-2000s~\cite{yang2006value,sentry,bussani2005trusted}. In 2012, GlobalPlatform and the Trusted Computing Group created standard specifications for TEEs. TEEs allow applications to use a secure region of memory (RAM) to store sensitive data and execute sensitive code without the risk of leakage or modification by unauthorized applications, including those with high privileges such as the OS and hypervisor. Today, TEEs are available from all major CPU vendors (e.g., ARM TrustZone~\cite{pinto2019demystifying}, Intel SGX (Software Guard Extensions)~\cite{cryptoeprint:sgx}, AMD SEV (Secure Encrypted Virtualization)~\cite{whitepaper:kaplan2016amd}).
        
        \subsubsection{Implementations comparison}
        Ménétrey et al~\cite{attestation-demystified} compared the TEE implementations of major vendors according to several features. We extract relevant comparison criteria in Table~\ref{table:tee-comparison}. It is noteworthy that other academic, non-commercial frameworks for TEEs exist, such as Keystone~\cite{keystone}, Sanctum~\cite{sanctum}, TIMBER-V~\cite{timber-v}, and LIRA-V~\cite{lira-v}, among others.

\begin{table}[!h]
\centering
\scriptsize
\renewcommand{\arraystretch}{1.5}  
    \begin{tabularx}{0.48\textwidth}{|X|X|X|X|} 
    \hline
    \multicolumn{1}{|c|}{\textbf{Features}} & \multicolumn{1}{|c|}{\textbf{SGX}} & \multicolumn{1}{|c|}{\textbf{TrustZone}} & \multicolumn{1}{|c|}{\textbf{SEV-SNP}}\\
    \hline
     \multicolumn{1}{|>{\centering\arraybackslash}X|}{Encryption} & \multicolumn{1}{|>{\centering\arraybackslash}X|}{\greencheck} & \multicolumn{1}{|>{\centering\arraybackslash}X|}{\redcross} &  \multicolumn{1}{|>{\centering\arraybackslash}X|}{\greencheck} \\
  \hline
     \multicolumn{1}{|>{\centering\arraybackslash}X|}{Integrity} & \multicolumn{1}{|>{\centering\arraybackslash}X|}{\greencheck} & \multicolumn{1}{|>{\centering\arraybackslash}X|}{\redcross} &  \multicolumn{1}{|>{\centering\arraybackslash}X|}{\graycheck} \\
   \hline
     \multicolumn{1}{|>{\centering\arraybackslash}X|}{Freshness} & \multicolumn{1}{|>{\centering\arraybackslash}X|}{\greencheck} & \multicolumn{1}{|>{\centering\arraybackslash}X|}{\redcross} &  \multicolumn{1}{|>{\centering\arraybackslash}X|}{\graycheck} \\
   \hline
     \multicolumn{1}{|>{\centering\arraybackslash}X|}{Local attestation} & \multicolumn{1}{|>{\centering\arraybackslash}X|}{\greencheck} & \multicolumn{1}{|>{\centering\arraybackslash}X|}{\redcross} &  \multicolumn{1}{|>{\centering\arraybackslash}X|}{\redcross} \\
   \hline
     \multicolumn{1}{|>{\centering\arraybackslash}X|}{Remote attestation} & \multicolumn{1}{|>{\centering\arraybackslash}X|}{\greencheck} & \multicolumn{1}{|>{\centering\arraybackslash}X|}{\graycheck} &  \multicolumn{1}{|>{\centering\arraybackslash}X|}{\greencheck} \\
   \hline
     \multicolumn{1}{|>{\centering\arraybackslash}X|}{Open source} & \multicolumn{1}{|>{\centering\arraybackslash}X|}{\graycheck} & \multicolumn{1}{|>{\centering\arraybackslash}X|}{\graycheck} &  \multicolumn{1}{|>{\centering\arraybackslash}X|}{\redcross} \\
    \hline
     \multicolumn{1}{|>{\centering\arraybackslash}X|}{Isolation level} & \multicolumn{1}{|>{\centering\arraybackslash}X|}{Intra-address space} & \multicolumn{1}{|>{\centering\arraybackslash}X|}{Secure World} &  \multicolumn{1}{|>{\centering\arraybackslash}X|}{Virtual Machine} \\
     
    \hline
    \end{tabularx}

   
\captionsetup{font=small}
\caption{Comparison of major TEEs implementations}
\label{table:tee-comparison}
\end{table}
        The considered criteria are:
        \textit{(1)} Encryption: DRAM of TEE instances is encrypted to ensure that no unauthorized access or memory snooping of the enclave occurs;
        \textit{(2)} Integrity: an active mechanism prevents DRAM of TEE instances from being tampered with;
        \textit{(3)} Freshness: protects DRAM of TEE instances against replay and rollback attacks;
        \textit{(4)} Local attestation: a TEE instance attests to another instance running on the same system;
        \textit{(5)} Remote attestation: a TEE instance attests to its genuineness to remote parties;
        \textit{(6)} Open-source: indicates whether the solution is publicly available;
        \textit{(7)} Isolation level: the level of granularity at which the TEE operates to provide isolation.
        (\greencheck) means full support of the feature; (\graycheck) means partial support; and (\redcross) means no support.\\
        \noindent\textbf{TrustZone.}
        TrustZone (TZ)~\cite{pinto2019demystifying} does not provide built-in remote attestation to establish the trustworthiness of the code loaded into the enclave. Instead, it relies on academic work~\cite{trustzone-ra-1,trustzone-ra-2,trustzone-ra-3,trustzone-ra-4} to add remote attestation support. Nevertheless, after enclave loading, TZ does not offer reliable mechanisms to protect against integrity and freshness attacks. It also logically isolates its enclave from the OS environment but does not encrypt its main memory, leaving the TZ enclave vulnerable to memory snooping. Additionally, TZ requires a side-OS to support the secure environment (\ie Secure World).\\
        \noindent\textbf{SEV-SNP.}
        SEV-SNP~\cite{whitepaper:kaplan2016amd} protects entire virtual machines (VMs), including their OS. However, it does not guarantee integrity or freshness against physical attacks. While AMD has provided documentation and specifications for SEV to assist software developers in understanding and implementing support for the technology, the actual implementation and firmware of AMD SEV are not open source.\\
        \noindent\textbf{SGX.}
        SGX~\cite{cryptoeprint:sgx} is the commercial TEE that supports the widest range of security features, particularly protection against physical attacks and encryption of secure memory. It also offers fine-grained security by enabling users to isolate specific parts of an application. SGX is further supported by extensive documentation and regular updates from Intel.
        \\

        Based on the previous comparison, SGX may be favored by developers and researchers for building TEE-based applications, including KVSs.

    \subsection{Intel SGX}
   Intel SGX~\cite{cryptoeprint:sgx} is a set of extensions to the Intel x86 CPU architecture for implementing a TEE. It aims to provide confidentiality and integrity for user-level sensitive data and computations, even when the underlying execution platform (including privileged software) is potentially malicious. SGX allows developers to split applications into untrusted (non-sensitive) parts that do not require security properties and trusted (sensitive) parts that do; the trusted parts run in a protected memory region called an \textit{enclave}.
   
   \subsubsection{Execution workflow.}
\begin{figure}
        \centering
        \includegraphics[width=0.42\textwidth]{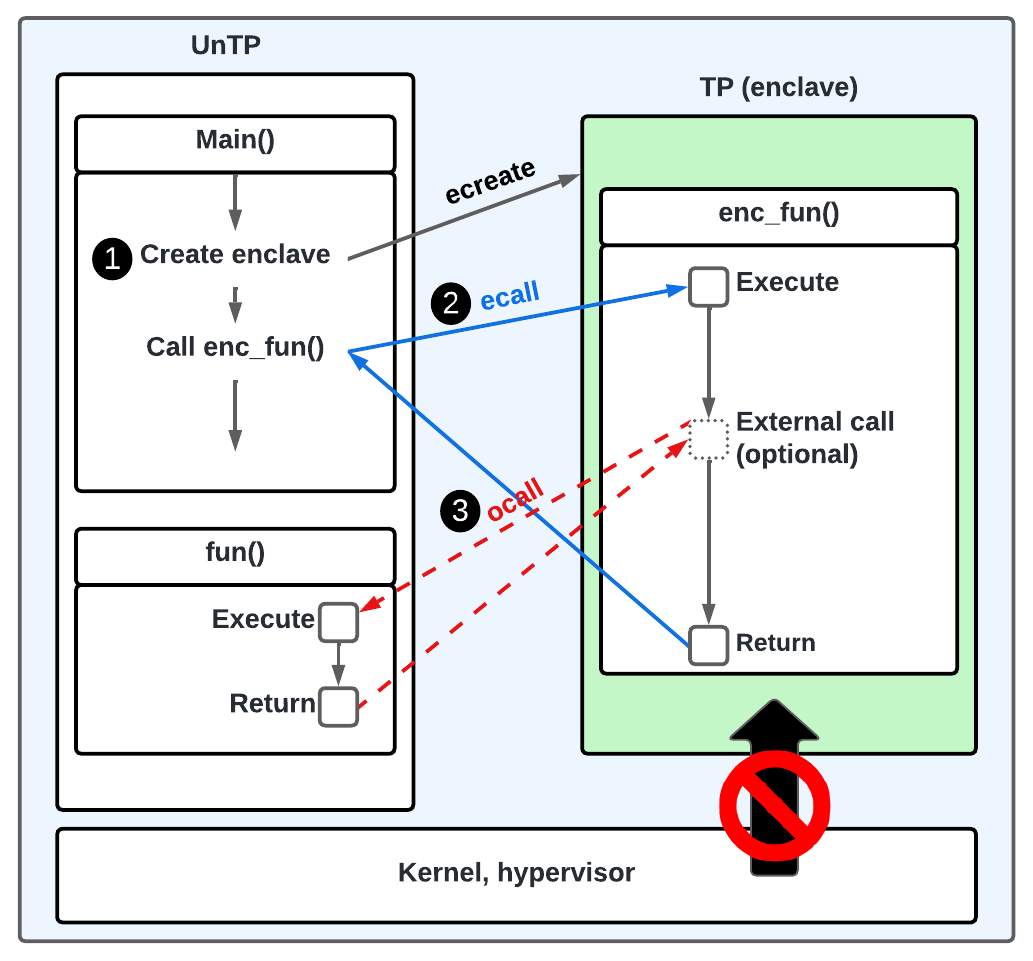}
        \caption{High level SGX execution workflow}
        \label{fig:sgx-execution-workflow}
    \end{figure}

    The execution workflow of SGX-based applications is described in Figure~\ref{fig:sgx-execution-workflow}. 
    An SGX application is subdivided into two parts: the Untrusted Part (UnTP), where the main program and routine functions are executed, and the Trusted Part (TP or enclave), where sensitive functions are executed.
    First, the \texttt{Main()} function of the UnTP runs until it reaches \circled{1} the \texttt{ecreate} instruction, which loads the enclave environment. The \texttt{Main()} function continues executing until it reaches \circled{2} an enclave call (\ie \texttt{ecall}) to an enclave function (\texttt{enc\_fun()}). At this point, the CPU changes its context and executes \texttt{enc\_fun()} until it returns. Optionally, during the execution of \texttt{enc\_fun()}, the CPU can \circled{3} issue a call to a function in the UnTP environment (\texttt{ocall}), such as a syscall. When the enclave function returns, the CPU restores its context and resumes the execution of \texttt{Main()}. 

       \subsubsection{Memory layout.} \label{sssec:memory-layout}
\begin{figure*}
            \centering
            \includegraphics[width=0.90\textwidth, height=0.18\textheight]{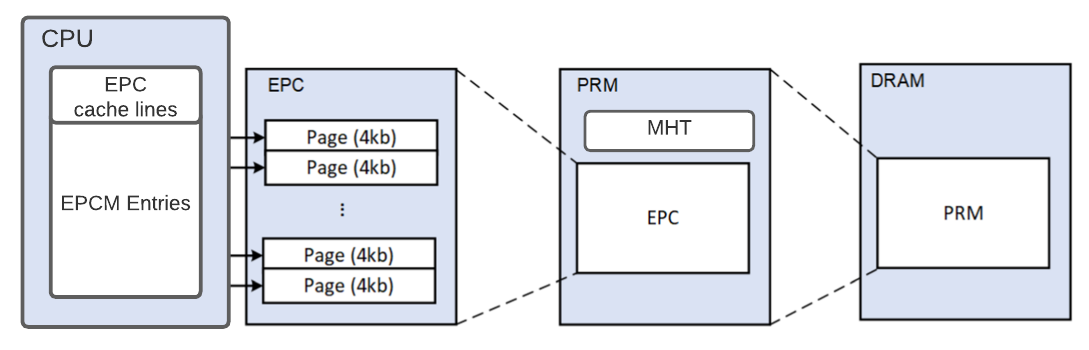}
            \caption{SGX Memory layout}
            \label{fig:sgx-memory-organization}
        \end{figure*}

       The memory organization of SGX is shown in Figure~\ref{fig:sgx-memory-organization}. 
        At boot time, a secure memory area called \textit{PRM} (Processor Reserved Memory), whose size is determined by the BIOS, is allocated as a subset of DRAM (Dynamic Random Access Memory). The PRM itself consists of two main parts: 
        \begin{itemize}[leftmargin=*] 
            \item \textit{Enclave Page Cache}: The \textit{EPC} is a reserved portion of system memory (DRAM) used to store the encrypted memory pages of the enclaves. These pages can only be decrypted at execution time by a key stored in the CPU, when they are in the physical processor core cache.
            When a trusted application exceeds the EPC size, the OS may evict some pages from the EPC to swap in other pages, following these steps: The CPU (1) reads the EPC page to be swapped out, (2) encrypts the contents of that page (second encryption), and (3) writes the encrypted page to regular, unprotected system memory. Since this process has inherent overhead, The more pages that are swapped out, the more performance drops.
            \item \textit{Merkle Hash Tree}: The \textit{MHT}~\cite{merkle-tree} maintains the Message Authentication Code (MAC) tags of the EPC pages in the form of a tree, where the leaves represent the individual MAC of each page in the enclave, and the root represents the summary MAC of all enclave content. 
        \end{itemize}
    
        \noindent To allow the creation of multiple enclaves, the EPC is divided into 4KB pages, each of which can be assigned to a different enclave instance. Since the system software responsible for allocating EPC pages (\eg OS kernel, hypervisor) is not fully trusted, SGX must verify the correctness of the allocation decisions. Therefore, SGX records the EPC allocation information in the \textit{EPCM} (Enclave Page Cache Map), a lookup table inside the CPU, where each entry corresponds to an allocated page.
    
        \subsubsection{Remote Attestation.} \label{ssec:remote-attestation}
        Remote attestation~\cite{web:sgx-remote-attestation} (RA) allows a remote client to verify that a particular application is running within an authentic (not forged) SGX enclave using an SGX-generated certificate. Remote attestation in SGX involves an architectural enclave called the Quoting Enclave (QE), provided by Intel, which generates a verifiable QUOTE to prove the authenticity of an enclave.
        Two RAs models are used : 
        \begin{itemize}[leftmargin=*]
            \item \textbf{EPID-based RA:} Enhanced Privacy ID (EPID)-based attestation~\cite{web:sgx-remote-attestation} is a cryptographic protocol that provides anonymous attestation. This method allows an enclave to generate a QUOTE while preserving privacy through the use of digital signatures and zero-knowledge proofs~\cite{EPID}. The QUOTE can then be verified online by the Intel Attestation Service (IAS).
            EPID-based RA ensures the authenticity and integrity of an enclave without revealing the identity of its signer. It is primarily used in scenarios where anonymity is crucial, such as peer-to-peer networks, distributed ledger technologies (\eg blockchain), or any situation where the identity of the attesting device needs protection. Despite its advantages, support for EPID-based RA is deprecated, and Intel is now focusing on the newer DCAP scheme.
            
            \item \textbf{\textit{DCAP}-based RA:} \textit{DCAP} (Data Center Attestation Primitives)-based attestation~\cite{web:sgx-remote-attestation,scarlata2018supporting} is a remote attestation scheme provided by Intel, specifically designed for data center environments. It enables the creation of on-premises attestation services, reducing the need for interactions with Intel's online services. This approach is particularly beneficial for cloud service providers (CSPs) who prefer to manage the attestation process in-house. DCAP allows CSPs to cache essential SGX certificates from Intel within their local infrastructure, thereby authenticating enclaves without relying on the IAS. Essentially, DCAP decentralizes the attestation protocol, facilitating faster enclave attestation.
        \end{itemize}
        
        \subsubsection{Sealing.} 
       Sealing~\cite{web:sgx-sealing} allows in-enclave data to be securely stored on persistent storage. The data is encrypted within the enclave using a secret sealing key derived from the \textit{Root Sealing Key} (a cryptographic key embedded in the CPU fuse array during manufacturing) before being stored outside the enclave. This encryption provides confidentiality and integrity assurances for the sealed data. Intel SGX offers two binding policies for sealing keys, based on two different measurements:
        \begin{itemize}[leftmargin=*]
            \item \textit{MRENCLAVE} stands for \textit{Enclave Measurement} and refers to a hash value of the enclave code and static data at the time of enclave creation.
            \item \textit{MRSIGNER} stands for \textit{Enclave Signer Measurement} and refers to a hash value of the identity of the entity that signed the enclave (typically an independent software vendor or a certificate authority). 
        \end{itemize}
        If the sealing key is tied to a specific MRENCLAVE, any change that affects the enclave's measurement will result in a different sealing key, making the sealed data undecipherable. In other words, the MRENCLAVE-based sealing is used to share data between enclave instances with the same MRENCLAVE.
        On the other hand, if the sealing key is bound to an MRSIGNER, the data can be unsealed by any enclave signed by the same signer as the one that sealed the data.

    \subsection{SGX performance}
    Some critical limitations directly impact the performance of applications running in SGX enclaves and hinder their adoption. Two primary sources of performance degradation should be considered when designing secure applications with SGX:
        \\
        \noindent \textbf{Page faults}.
        If the secure portion of the application exceeds the size of the EPC, the extra pages will be encrypted and swapped out of the EPC memory region. Accessing these pages will trigger costly decryption and verification checks when they are loaded back into the EPC. Therefore, it is advisable to wisely select the components (both code and data) to be secured inside the enclave to avoid expensive page swapping.\\
        \noindent \textbf{Environment context switch}.
        Environment context switch refers to switching between the untrusted part of an application and the trusted part in an SGX enclave. A context switch is triggered by \texttt{ecalls} and \texttt{ocalls} and results in significant performance degradation, being 5.5 times more expensive than a user-mode context switch~\cite{avocado,sgx-hotcalls}. Excessive environment context switches can be caused by an application design that frequently alternates between trusted and untrusted parts, or by an excessive number of in-enclave system calls (syscalls).

    \subsection{SGX versions}
    Intel SGX CPUs are available in two flavours: 
    \noindent\textbf{Client SGX.}
    Client SGX systems provide comprehensive protection against software and hardware (physical) confidentiality and integrity attacks. To offer integrity protection, client SGX relies on the MHT at the expense of maximum secure memory, which does not exceed 128 MB. Any additional secure pages will trigger page swapping on Linux systems. An evolution of client SGX, namely SGX2, introduces EDMM (Enclave Dynamic Memory Management) feature. It allows an enclave to add more secure memory after it has already been loaded.\\
    \noindent\textbf{Scalable SGX.}
    Built on top of SGX2, Scalable SGX was released in 2021, focusing on server-grade processors. Scalable SGX significantly increases the amount of secure memory available to the enclave to 512GB. It also introduces support for multi-socket CPUs. Furthermore, Intel switched from Memory Encryption Engine (MEE) to Total Memory Encryption -- Multi-Key (TME-MK)~\cite{web:tme-mk} for faster memory encryption and execution. However, the increase in secure memory and performance compromises the security guarantees of scalable SGX CPUs, which sacrifice MHT and thus protection against hardware-based integrity attacks.

\section{TEE-based KVSs threat model} \label{sec:threat-model}
    We surveyed many TEE-based KVSs~\cite{veritasDB,shieldstore,speicher,precursor,tweezer,concerto,elsm,avocado,treaty,engraft-tiks} to understand the threat models they use. A common factor among these systems is their reliance on Client SGX as the TEE implementation. The likely motivation behind this choice, despite the fact that some KVSs appeared before scalable SGX, is that \textit{"among the commercially available TEEs, Client SGX is the only one that provides integrity guarantees against a physical attacker"}~\cite{sgx-integrity}, despite its limitation of having a small secure memory (128MB). Consequently, we detail the Client SGX threat model\cite{cryptoeprint:sgx} that is considered by SGX-based KVS nodes.\\

    SGX-based KVSs are designed to mitigate the following threats inside the enclave (EPC):
    \begin{itemize}[leftmargin=*]
    
        \item \textbf{Malicious Software:} EPC is protected against malware that may attempt to access sensitive data (KV records) or tamper with application workflow running in EPC.
        
        \item \textbf{Insider Attacks:} EPC is protected against attacks from users or processes with legitimate access to the system (including OS, hypervisors) but who may try to abuse that access to steal sensitive information (KV records).
        
        \item \textbf{Physical Attacks:} SGX helps mitigate the risk of attacks on the system's hardware components by ensuring that sensitive data (KV records) inside EPC is encrypted by hardware key and isolated, even from privileged software running on the system.

        \item \textbf{Alteration of code and static data:} SGX allows remote parties to ensure that the expected code and static data are loaded in enclave through RA.
        
    \end{itemize}

Lone SGX is not designed to handle SCAs, therefore the literature provides mitigation techniques on top of SGX. The surveyed TEE-based KVSs do not employ any mitigation techniques against SCAs, as these are notoriously costly and conflict with the objectives of the studied KVSs, which aim to maintain practical performance. However, some TEE-based SQL databases~\cite{opaque,oblidb,speed-sgx-encrypted-search} do leverage SCA mitigation techniques that may also be applicable to KVSs. Section~\ref{sec:sca} provides an overview of SCAs.

\section{TEE-based KVSs modules} \label{sec:kvs-modules}
\begin{figure}
        \centering
        \includegraphics[width=0.48\textwidth]{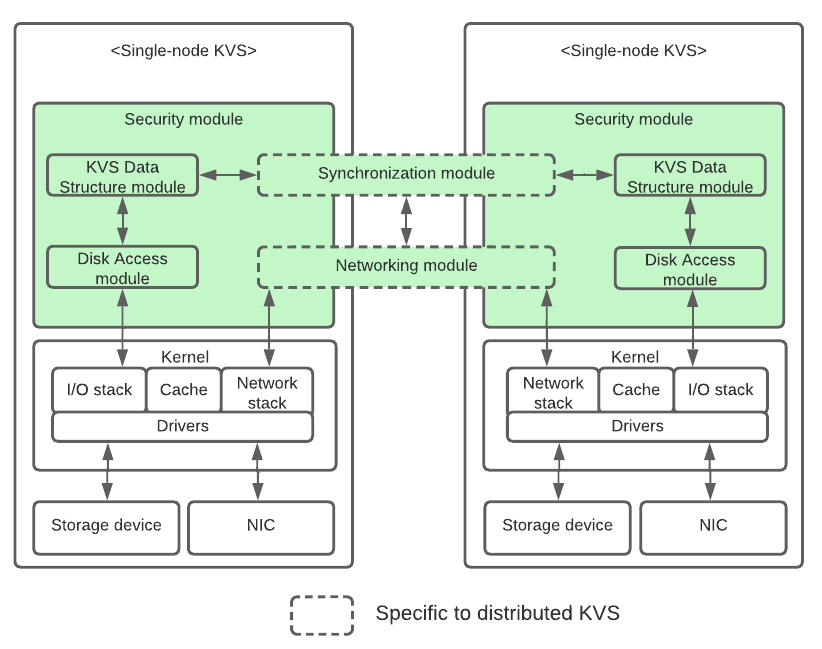}
        \caption{TEE-based Key-value stores generic architecture}
        \label{fig:KVS-architecture}
    \end{figure}

    TEE-based KVSs have a typical architecture that encompasses several modules. They are highlighted with green rectangles in Figure \ref{fig:KVS-architecture}. Rectangles with dashed lines represent modules that are specific to distributed KVS. \\
    \noindent\textbf{KVS' Data Structure module.} 
    The choice of the underlying data structure to store the KV pairs directly impacts the time complexity of the KVS for processing requests and its space complexity. The former point is independent of the use of Intel SGX, while the latter directly influences performance due to the limited size of secure memory. 
    The goal of this module is to survey the data structures used in building SGX-based KVSs and describe how they were adapted to fit the limited secure memory constraints of SGX. \\
    \noindent\textbf{Disk Access Optimization module.} This module encompasses a set of optimizations to efficiently access the storage device, through the enclave memory, without experiencing SGX-typical context switching overhead. It encompasses asynchronous storage syscall and the use of Storage Performance Development Kit (SPDK)~\cite{spdk}.\\
    \noindent\textbf{Networking Optimization module.} On the same way than Disk Access Optimization module, Networking Optimzation module encompasses a set of optimizations to efficiently access the network device (NIC) and process network packets. It encompasses asynchronous networking syscall and the use of Data Plane Development Kit (DPDK)~\cite{dpdk}. \\
    \noindent\textbf{Synchronization module.} 
    Multi-node (distributed) KVSs require communication to ensure data consistency and integrity across multiple nodes, using synchronization protocols that handle various system and networking errors (packet dropping, delays, alterations, \etc). 
    This module presents the synchronization protocols adopted by distributed SGX-based KVSs to ensure consistency across their nodes in a byzantine environment~\cite{pBFT} and how SGX benefits to them.\\
    \noindent\textbf{Security module.} 
    This module is a transversal component linked to all the previous modules to ensure they are correctly secured with the appropriate confidentiality and integrity guarantees, even outside the enclave. It encompasses a set of measures taken to extend the security of the enclave to sensitive data located outside the enclave, whether due to the lack of secure memory space or because of persistency (data exits the enclave boundaries when it is in storage disk). \\
    
    The remaining of this section details the different building blocks and/or implementations of each module.
     
    \subsection{KVS' data structure module} \label{ssec:data-strcuture}
    The surveyed KVSs use the following data structures: Hash table, Skip list and Log-Structured Merge Tree (LSMT). We systematically explain the architecture of each of them, their space complexity, \textit{get/put} operation workflow and their adaptation for SGX considering its small secure memory.
        
        \subsubsection{Hash table}  
        A hash Table~\cite{hash-table} is a data structure that implements an associative array abstract data type, a structure that can map keys to values. It uses a hash function to compute an index into an array of buckets (or slots), from which the desired value can be found. A hash function takes an input (or 'key') and returns an integer, which is typically the index where the KV pair is stored in the hash table. The goal is to distribute keys uniformly across the buckets. Each bucket contains a list (or chain) of KV pairs that mapped to the same bucket.\\
        \noindent \textbf{Space complexity.} 
        Considering a hash table with \textit{M} buckets and \textit{N} KV pairs, the space complexity of such structures is \textit{O(M+N)}.
        \\
        \noindent \textbf{Put operation.} 
        To add a KV pair to the hash table, (i) we compute the hash value of the key using the hash function, (2) we use the hash value to determine the index in the array, and (3) we insert the KV pair into the appropriate bucket. If the buckets are uniformly filled, this \textit{put} operation results in an \textit{O(1)} complexity. In the worst case, if all keys hash to the same index (poor hash function or high load factor), all KV pairs (\textit{N}) end up chained in the same bucket. In this scenario, the insertion requires traversing a list of length \textit{N}, resulting in an \textit{O(N)} complexity.
        \\
        \noindent \textbf{Get operation.} 
        The \textit{get} operation follows the same pattern as \textit{put}, where we use the hash function on the key to locate the appropriate bucket and search for the element within it. Consequently, depending on the distribution of KV pairs across the buckets, the complexity can be \textit{O(1)} in the best case and \textit{O(N)} in the worst case.
        \\
        \noindent \textbf{Adaptation to SGX.} 
        ShieldStore~\cite{shieldstore} is an SGX-based KVS that uses a hash table to store its KV pairs. To save SGX memory, the entire data structure is kept outside the enclave. For confidentiality, each KV pair in each bucket is encrypted inside the enclave using an in-enclave encryption key before being stored outside. For integrity, a Merkle tree over the buckets is maintained outside the enclave, with its root secured inside the enclave.
        \\


        \subsubsection{Skip list} 
        A skip list~\cite{skiplist} is an in-memory probabilistic data structure that allows fast search, insertion, and deletion operations within an ordered sequence of elements (according to the key part of KV record). Skip lists are an alternative to balanced trees and provide a simpler and often more efficient implementation for certain types of data management tasks. It consists of multiple layers of linked lists. 
        Skip list also introduces the concept of probabilistic balancing by using randomization to maintain an efficient structure (\ie nodes are promoted to higher levels based on a probabilistic criterion).
        Skip list leverages the following key principles:
        \begin{itemize}[leftmargin=*]
            \item \textit{Nodes:} Each node in a skip list contains a data (\ie KV pair) and a set of forward pointers. The number of forward pointers (levels) a node has is determined randomly when the node is inserted.
            \item \textit{Levels:} The skip list has multiple levels, with the bottom level containing all elements. Higher levels contain fewer elements, creating shortcuts for faster traversal.\\
        \end{itemize}
        \noindent\textbf{Space complexity.}
        Every element in the skip list appears in the base level (level 0), which is a simple linked list. This contributes \textit{O(N)} space, where \textit{N} is the number of elements. Each element has a probability \textit{p} of being promoted to the next level. Commonly, \textit{p} is set to \textit{0.5}, meaning each level has about half as many elements as the level below it. The expected number of levels in a skip list is \textit{O(log(N))}. This is because, with probability \textit{p}, a node appears in level 1; with probability $p^2$, it appears in level 2, and so on. Thus, the total number of nodes is modeled by a geometric series of ratio \textit{p=0.5}, leading to a total number of nodes bounded by \textit{2N}. Thus, the average space complexity of skip list, without considering the pointers, is \textit{O(N)}.
        \\
        \noindent\textbf{Get operation.}
        A skip list is composed on average of \textit{O(log(N))} levels as we explained above. During a search, we start from the highest level and move forward until the next node's key exceeds the target key, then drop down a level. This process ensures that we are practically going through the levels, leading to an average time complexity of \textit{O(log(N))}.
        \\
        \noindent\textbf{Put operation.} 
        (i) Inserting or updating an element requires finding the appropriate position, which on average takes \textit{O(log(N))} due to the search process explained above. (ii) Once the position is found, the new element is inserted into the multiple levels based on random level assignment. Since the expected number of levels for any node is \textit{O(log(N))}, the insertion process also averages to \textit{O(log(N))}.
        \\
        \noindent\textbf{Adaptation to SGX.} 
        Avocado~\cite{avocado} is a distributed SGX-based KVS where each nodes utilizes a skip list for indexing its underlying data structure. The skip list index, along with the keys, pointers to values, and their individual MACs, are stored within the enclave. The values are stored outside the enclave, in an encrypted form, using an in-enclave encryption key. 
        Overall, this design attempts to ensure a balance between security (\ie by protecting keys, cryptographic keys and integrity checks inside the enclave) and performance (\ie by storing the bulkier values outside the enclave to avoid overloading EPC).
        \\
        

        \subsubsection{Log Structured Merge Tree. } \label{ssec:data-structure-lsmt}
       LSMT~\cite{lsmt} is a disk-based data structure designed for high-throughput write operations and efficient read operations in systems that handle large volumes of data and that require persistence. It is widely used in modern persistent databases and storage systems like RocksDB~\cite{web:rocksdb}, and Speicher~\cite{speicher}. Behind LSMT name, two key concepts are leveraged:
       (1) Log-structured: Data is initially written to a log in an append-only manner to enforce sequential writes, which are faster than random writes, especially on disk storage;
       and (2) Merge tree: Data is organized in multiple levels or tiers, each level consisting of sorted runs of data. Periodically, data from smaller, more frequently accessed levels is merged into larger, less frequently accessed levels.\\
       LSMTs leverage three key principles:
       \begin{itemize}[leftmargin=*]
       
        \item \textit{MemTable:} The LSMT begins with an in-memory table called the MemTable. All writes are first recorded here. Once the MemTable reaches a certain size, it is flushed to disk, creating a new sorted file called an SSTable (Sorted String Table).
        
        \item \textit{SSTable:} SSTables are immutable and stored on disk. Each SSTable is organized in a set of blocks (tailored to the same block size as the disk), each block contains a set of KV pairs and the whole SSTable is sorted by key. When a MemTable is flushed, it becomes a new SSTable in Level 0.
        
        \item \textit{Compaction:} To manage the growing number of SSTables and maintain read efficiency, compaction processes merge SSTables, invloving merging overlapping SSTables within the same level or between adjacent levels, deleting obsolete versions of keys and consolidating fragmented data. The process creates a new bigger SSTable at highest level. Compaction helps in maintaining the sorted order and removing stale data, thus optimizing read performance.
        
       \end{itemize}
        
       \noindent\textbf{Space complexity.} LSMT spans both main memory and disk storage. However, since enclave secure memory only encompasses main memory, we focus on the spatial complexity of the MemTable. The complexity depends on the data structure used to implement the MemTable. For classical architectures (\eg B-tree, skip list, etc.), the spatial complexity of the MemTable is \textit{O(N)}, where \textit{N} represents the maximum number of records a MemTable can store before flushing.
       \\
       \noindent\textbf{Put operation.} 
       The temporal complexity of \textit{put} operation is amortized to \textit{O(1)} as explained in the following: 
       (i) Put operations are first appended in a write-ahead log (WAL) to ensure durability, which is an \textit{O(1)} operation.
       (ii) Data is then written to the MemTable. Assuming that MemTable is a balanced data structure that record \textit{N} KV pairs before flushing, it translates to an \textit{O(log(N))} operation. But considering the amortized cost over multiple operations and the efficient management of writes, it's often treated as \textit{O(1)}.
       (iii) When the MemTable is full, it is flushed to disk as an SSTable. Similarly to the previous operation, it is amortized to \textit{O(1)} considering that flushing spreads the cost over many writes.
       (iv) Finally, if compaction is triggered, its cost is also spread over many writes, leading to an amortized complexity of \textit{O(1)} per write.
        \\
       \noindent\textbf{Get operation.}
       (i) Read operations first check the MemTable, which takes \textit{O(log(N))} if the MemTable is implemented as balanced tree, assuming that \textit{N} is the number of entries of the MemTable. 
       (ii) If the data is not found in the MemTable, the system searches through the SSTables, starting from the most recent and moving to older ones. To spead up this search process, bloom filters and other indexing structures are often used to quickly eliminating SSTables that do not contain the queried key. Assuming that the previous optimization techniques allow us to search only one SSTable per level and that \textit{M,K} are respectively the number of entries in a single SSTable and the number of SSTable levels, the complexity of searching in SSTables is \textit{O(K.log(M))}.
       The summarized complexity of \textit{get} operation in LSMT is \textit{O(log(N)+K.log(M))}.
        \\
        \noindent\textbf{Adaptation to SGX.}
        Speicher~\cite{speicher} is a pioneering SGX-based KVS that leverages LSMT as its underlying data structure. For the MemTable implementation, it uses a skip list, similar to Avocado~\cite{avocado}. Consequently, it inherits the same partitioning design between enclave and non-enclave as described in \textit{\ref{ssec:data-structure-lsmt}: Adaptation to SGX}. Before persisting an SSTable, each block is encrypted using an in-enclave encryption key for confidentiality. Additionally, a Merkle tree is built over the SSTables of each level, where their root are kept in enclave, ensuring SSTables integrity and freshness. 
        \\


    \subsection{Disk Access Optimization module}\label{sec:disk-access-module}
       One of main source of overhead of SGX applications comes from context switching between the trusted and untrusted environments. In the context of SGX-based KVS, this switch is typically triggered by synchronous storage syscall triggered within an enclave, usually for persisting in-enclave data on disk. Two approaches are used to overcome this overhead:

        \subsubsection{Asynchronous storage syscall} 
        Asynchronous storage syscall allows to execute storage operations asynchronously by leveraging a producer/consumer model. Syscall is issued from the enclave thread by placing a request into the producer queue (request queue). Then, a non-enclave thread, continuously probing the queue, processes these requests. When the syscall returns, the non-enclave thread places the result into the consumer (response) queue to be used by the enclave thread.
        Thus, enclave thread does not exit the enclave environment and consequently does not perform the expensive context switch. Shielded execution frameworks such as Scone~\cite{scone} and Eleos~\cite{eleos} provide asynchronous storage interface.

        \subsubsection{Storage Performance Development Kit}
        SPDK~\cite{spdk,web:spdk} is a collection of tools and libraries designed to significantly improve the performance of storage applications by utilizing user-space, polled-mode drivers, and minimizing CPU overhead. SPDK achieves this by bypassing the traditional kernel storage stack, thus reducing latency and improving throughput.
        SPDK is built around several key principles:
        
        \begin{itemize}[leftmargin=*]
        \item \textbf{User space drivers:}
        Virtual memory is typically divided into kernel space and user space, with storage drivers typically operating in kernel space. However, SPDK's drivers run in user space while still directly controlling storage hardware.
        To utilize SPDK, the OS must first release control (unbind) of the device. On Linux, this is achieved through \textit{sysfs}. SPDK then binds the device to special dummy drivers like vfio, preventing the OS from reclaiming the device. Subsequently, SPDK replaces the OS storage stack with its own implementations in C libraries, which include block device abstraction, block allocators, and filesystem components.

        \item \textbf{Polled mode drivers:}
        Unlike interrupt-driven drivers, SPDK employs polled-mode drivers. Specifically, when a user provides a callback function for an I/O operation, SPDK continuously polls the device to check for the completion of that I/O operation. When the I/O operation finishes, SPDK triggers the provided callback function. This polling mechanism ensures low-latency and high-performance storage operations by avoiding the overhead associated with interrupts and context switches.
    
        \item \textbf{Zero-copy data path:}
        In traditional I/O operations, data typically has to be copied multiple times. For instance, in order to write to storage device, data need to be copied from the user application buffers to kernel staging buffers, then from kernel staging buffers to storage device. These additional copies introduce latency and consume CPU resources.
        Zero-copy allows user applications to transfer data directly from the host memory to storage without using any staging buffer inside the kernel.
 
        \item \textbf{Modular Design:}
        SPDK is modular, enabling developers to use only the components necessary for their specific use case. This is particularly useful in SGX context where the size of secure memory is limited. 
        
        \end{itemize}

\begin{figure*}
                \centering
                \subfigure[Combining SPDK and DMA optimizes disk access within SGX enclaves by using the SPDK library to provide NVMe driver access in user space while the DMA controller handles data transfer. The combination eliminates dependence on the kernel storage driver, kernel storage stack, and kernel buffers. Nevertheless, SPDK's buffers should reside outside the enclave to be registered by DMA controller and to avoid costly EPC paging.\label{fig:spdk_dma}]{
                    \includegraphics[width=0.29\textwidth]{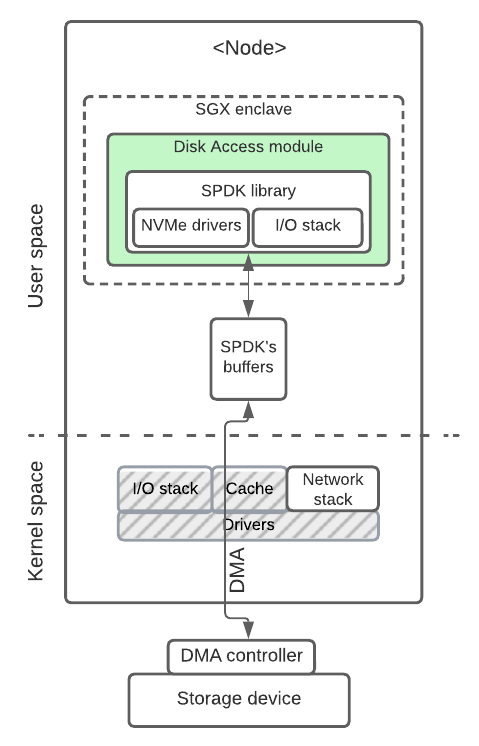}   
                }
                \hspace{0.05\textwidth}%
                \subfigure[Combining DPDK and RDMA optimizes data exchange within SGX enclaves by using the DPDK library to provide NIC driver access in user space while RDMA transfers data between two remote devices. This eliminates dependence on the kernel's NIC drivers, network stack, and cache. Nevertheless, DPDK's buffers should reside outside the enclave to be registered by DMA controller.\label{fig:dpdk_rdma}]{
                    \includegraphics[width=0.55\textwidth]{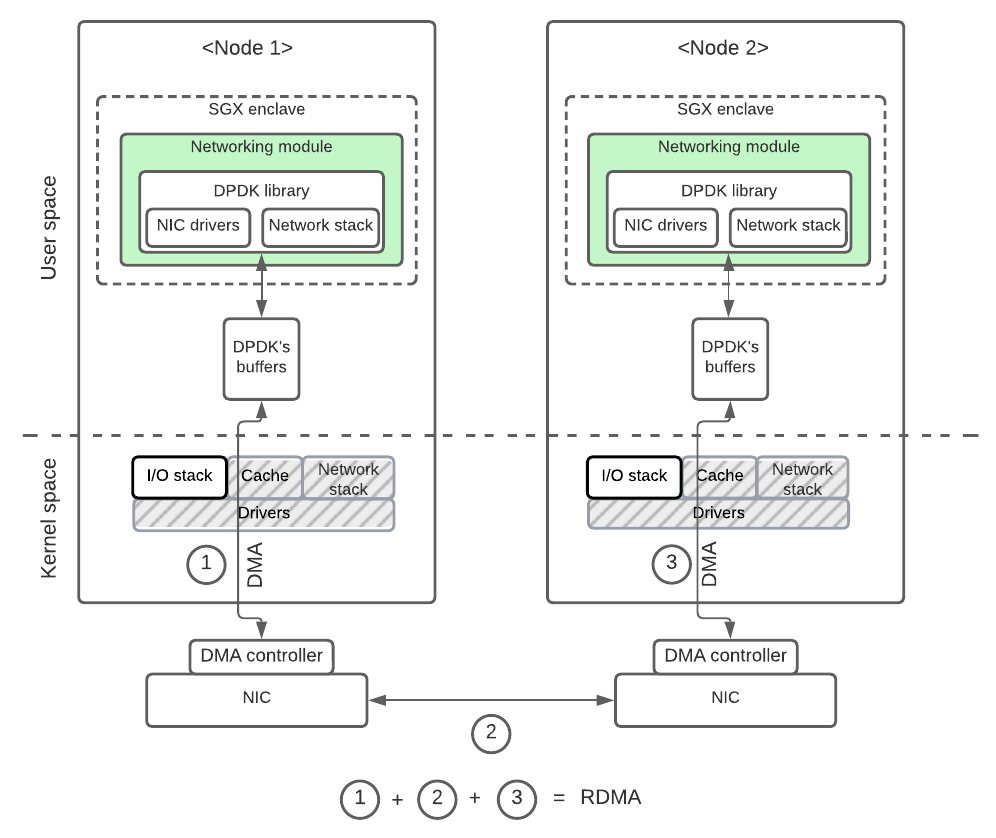}
                }
                \caption{Direct storage and direct transfer technologies in SGX-based KVS} 
            \end{figure*}

        \noindent\textbf{Coupling with DMA.}
        DMA (Direct Memory Access) is a feature of computer systems that allows certain hardware subsystems (\eg storage devices) to transfer data to or from memory without involving the CPU, thus freeing up CPU resources for other tasks. The data transfer is supervised by the DMAC (DMA Controller), which only involves the CPU for initialization (\ie setting the source and destination addresses, amount of data to be transferred, and direction of transfer). DMA is often coupled with SPDK to manage data transfer from userspace between storage device and user application buffers without intermediate copying (\ie zero-copy) and without involving the CPU (\ie using DMAC). 
        \\ 
        \noindent\textbf{Adaptation to SGX.}
         In the context of Intel SGX, SPDK's user space drivers can be installed within the enclave to enable enclave threads to issue I/O storage operations without exiting their enclave environment. Speicher~\cite{speicher} is an SGX-based KVSs that leverage in-enclave SPDK logic to improve I/O storage performance. However, to avoid expensive EPC paging, it hosts SPDK buffers outside the enclave. Figure~\ref{fig:spdk_dma} shows the combination between SPDK and DMA in the context of SGX applications and the bypassed kernel components.
         Performance measurements of Speicher's I/O throughput in GB/S (Gigabytes per second) under different block size conditions of SSTables reveal that the throughput of Speicher with adapted in-enclave SPDK is similar to that of RocksDB~\cite{web:rocksdb} with native -without enclave- SPDK (1.25GB/s for blocks of 16KB for instance), indicating that I/O-related overhead for enclaves are successfully mitigated.

    \subsection{Networking Optimization module} \label{sec:networking-module}
    
    Another reason of context switch between the trusted and untrusted environment in SGX applications, alongside accessing to storage device, is accessing to network device (NIC, \ie Network Interface Card) to transfer packets through the network. 
    Thus, Networking Optimization module is specific to distributed KVS where high-performance networking is essential for supporting low-latency communication between nodes. Traditional inter-node communication involves using synchronous networking syscall based on sockets to process network packets. Similar to persisting data, processing network packets force enclave thread to leave its security context, leading to a non-negligible overhead. TWo approaches are used to overcome this issue: 
    
    \subsubsection{Asynchronous transfer syscall} 
    Like asynchronous storage syscall, it is an exitless syscall approach for inter-node data transfer based on sockets, where networking I/O operations do not block or prevent the execution of other code. Specifically, enclave thread delegates the syscall of packet transfer from/to NIC (Network Interface Card) to an external thread while the enclave thread continues to execute its code. The purpose is to avoid the context switch induced by in-enclave syscalls and optimize the CPU usage time. 
    
    \subsubsection{Data Plane Development Kit} 
    DPDK~\cite{dpdk,web:dpdk} is a set of libraries and drivers for fast packet processing, primarily used in network applications. It allows developers to build high-performance networking applications by bypassing the kernel network stack and accessing the hardware directly. Similar to SPDK, DPDK provides user-space drivers for various NICs and polling mode drivers. It also leverages Hugepages as described below.\\
    \noindent\textbf{Hugepages.}
    In most OSs, memory is managed in small chunks of 4KB pages. Hugepages are much larger memory pages, typically 2 MB or even 1 GB in size, depending on the hardware and OS configuration. Using hugepages in DPDK context allows for efficient memory allocation by providing large contiguous blocks of memory to handle high-speed packet processing while reducing fragmentation. Also, by reducing the number of pages (by increasing the size of individual pages), hugepages decrease the number of page table entries and TLB misses. This translates to lower latency and higher throughput in packet processing tasks.
    \\
    \noindent\textbf{Coupling with RDMA.}
    If DMA is a feature for transferring data between local peripheral devices and main memory without involving the CPU, RDMA (Remote DMA) extends this capability to transfer data between the main memory of two remote systems over a network, without involving the CPUs of either system. RDMA is often coupled with DPDK to manage packet transfers between two remote userspace memories, while bypassing the CPU.
    \\
    \noindent\textbf{Adaptation to SGX.}
     In the context of Intel SGX, DPDK's user space drivers can be installed within the enclave to enable enclave threads to issue I/O networking operations without exiting their enclave environment. Avocado~\cite{avocado} and Treaty~\cite{treaty} are both distributed SGX-based KVSs that use in-enclave DPDK logic to improve networking performance. However, to avoid costly EPC paging, Hugepages that host DPDK buffers are instantiated outside the enclave.
     Figure~\ref{fig:dpdk_rdma} shows the combination of DPDK and RDMA in the context of SGX applications and the bypassed kernel components. Performance measurements on Avocado show that its network stack based on DPDK is 1.66x faster in processing packets than sockets-Scone, an in-enclave and asynchronous secure socket by Scone~\cite{scone} that does not leverage DPDK optimizations.
     \\

    Table~\ref{tab:bypassed-components} summarizes the kernel components that are bypassed for each storage/networking optimization technology as well as their ability to bypass the CPU.

\begin{table*}
\centering
\renewcommand{\arraystretch}{1.5}  
\scriptsize
    \begin{tabularx}{\textwidth}{X|X|X|X|X|X|X|} 
    \cline{2-7}
    \multicolumn{1}{c}{}  & \multicolumn{5}{|c|}{\textbf{Bypassed kernel components}} & \multirow{2}{*}{\textbf{\makecell{CPU\\ bypass}}}\\
    \cline{2-6}
    
    & \multicolumn{1}{c|}{\textbf{Storage stack}} & \multicolumn{1}{c|}{\textbf{Storage driver}} & \multicolumn{1}{c|}{\textbf{Network stack}} & \multicolumn{1}{c|}{\textbf{\makecell{NIC\\ driver}}} & \multicolumn{1}{c|}{\textbf{\makecell{Cache\\(buffers)}}} &  \\
    \hline

    \multicolumn{1}{|c|}{\textbf{SPDK+DMA}} & \multicolumn{1}{c|}{\greencheck} & \multicolumn{1}{c|}{\greencheck} & \multicolumn{1}{c|}{\redcross} & \multicolumn{1}{c|}{\redcross} & \multicolumn{1}{c|}{\greencheck} & \multicolumn{1}{c|}{\makecell{\greencheck\\ (w DMA)}}\\
    \hline

    \multicolumn{1}{|c|}{\textbf{DPDK+RDMA}} & \multicolumn{1}{c|}{\redcross} & \multicolumn{1}{c|}{\redcross} & \multicolumn{1}{c|}{\greencheck} & \multicolumn{1}{c|}{\greencheck} & \multicolumn{1}{c|}{\greencheck} & \multicolumn{1}{c|}{\makecell{\greencheck\\ (w DMA)}} \\
    \hline

    \end{tabularx}

\captionsetup{font=small} 
\caption{Bypassed kernel components w.r.t direct storage and direct transfer technologies}
\label{tab:bypassed-components}
\end{table*}

    \subsection{Synchronization module} \label{ssec:synchronization-module}
        The surveyed distributed SGX-based KVSs leverage three different synchronization protocols, implemented inside the enclave. This approach ensures reliable execution of synchronization code, allowing developers to avoid addressing the full spectrum of Byzantine failures (as handled by Byzantine Fault Tolerance - BFT - protocols~\cite{pBFT}). Indeed, since enclaves are designed to prevent misbehavior, the synchronization module in SGX only needs to address crash and networking failures, which do not result from arbitrary cheating. Consequently, all employed synchronization algorithms in distributed SGX-based KVSs are CFT (Crash Fault Tolerant) rather than BFT (Byzantine Fault Tolerant). 
        In a nutshell, using CFT logic in SGX enables systems to be resilient against the same malicious behaviors considered in BFT, while requiring only $2f+1$ nodes, as in CFT systems, instead of $3f+1$ (\textit{f} being the number of faulty nodes).\\

        Synchronization module is subdivided into two classes: Consensus protocols for non-transactional KVSs~\cite{avocado,engraft-tiks} (Raft~\cite{raft} and ABD~\cite{ABD}) and a consensus protocol for transactional KVSs~\cite{treaty} (Two-phase commit~\cite{2pc}). We define each protocol and explain for Raft and ABD how the write and read operations are handled through the nodes. For 2PC, we explain how the transactions are distributively handled across the participants.

        \subsubsection{Raft}
        Raft~\cite{raft} is a distributed consensus algorithm designed as an alternative to the Paxos~\cite{paxos} algorithm. It is used for managing a replicated log in a distributed system to ensure crash tolerance and consistency. Specifically, in  a system of \textit{n} replicated nodes, Raft can tolerate $f\le[\frac{n-1}{2}]$ crashes.  
        In Raft, the system is divided into three roles: leader, follower, and candidate. These roles are determined through an election process. The leader is responsible for managing the replication of the log and handling client requests. Followers replicate the log and respond to client requests forwarded by the leader. Candidates are nodes attempting to become the leader by initiating leader election when no current leader is detected.
        Raft operates through a series of terms, each with its own unique leader. During each term, a leader is elected through an election process in which nodes exchange messages to determine the leader. Once a leader is elected, it maintains its position until it fails or another node with a higher term is elected.\\
        \noindent \textbf{Write operation.}
        The core idea of Raft is log replication. The leader receives client requests, appends them to its log, and then replicates the log entries to followers. Once a majority of followers have confirmed receiving an entry, it's considered committed, and the leader can apply the entry to its state machine and respond to the client.\\
        \noindent \textbf{Read operation.}
        In Raft, read operations are typically handled by the leader. When a client sends a read request to any node in the system, that node forwards the request to the current leader. The leader then responds to the read request by accessing its own state machine and returning the requested data to the client. This operation is performed by the leader to ensure that the client receives the most up-to-date data.\\`

        \subsubsection{ABD}
        The ABD consensus protocol~\cite{ABD} is another distributed consensus algorithm, specifically designed for systems where read operations are more frequent than write operations. It is a lightweight protocol that optimizes for read operations while ensuring consistency across the distributed system. Similar to Raft, in a system of \textit{n} nodes, ABD can tolerate $f\le[\frac{n-1}{2}]$ crashes. 
        In the ABD protocol, each node maintains a data structure containing two timestamps: a read timestamp and a write timestamp. These timestamps are used to track the most recent read and write operations performed on a particular data item. \\
        \noindent \textbf{Write operation.}
        When a node wants to perform a write operation on a data item, it initiates a write process by assigning a new timestamp to the data item. This timestamp is higher than any previous timestamps associated with the data item. The node then propagates the write request to other nodes in the system. Upon receiving a write request, a node compares the timestamp associated with the request with its own timestamps for the data item. If the request timestamp is greater, the node updates its own timestamp and acknowledges the write operation. If the request timestamp is lower or equal, the node ignores the request, ensuring that only the most recent write operation is accepted.\\
        \noindent \textbf{Read operation.}
        In contrast to Raft, any node can handle read operation itself without forwarding the task to a unique leader. Still, to ensure that a read operation returns the most recent value and maintains consistency across the distributed system, the ABD protocol requires that a read operation contacts a quorum of nodes. This ensures that the read operation encounters the most recent write operation during its execution.\\

        \subsubsection{Two-phase commit}
        The Two-Phase Commit (2PC)~\cite{2pc} protocol is a distributed algorithm crucial for maintaining atomicity in distributed transactions, a key property related to both consistency and integrity in the context of transactions. It orchestrates a coordinated decision-making process among all participating nodes. 2PC leverages two sequential phases.\\
        \noindent \textbf{Voting phase.}
        The coordinator node solicits votes from all participants regarding their readiness to commit the transaction (prepare request). Participants respond with either a ``Yes'' or a ``No'' based on their ability to commit. If all participants agree to commit, the coordinator proceeds to the second phase. Otherwise, if any participant votes ``No'' or fails to respond within a timeout period, the coordinator decides to abort the transaction.\\
        \noindent \textbf{Decision phase.}
        The coordinator instructs all participants to commit the transaction (commit request). Upon receiving the commit request, each participant commits the transaction and acknowledges the coordinator. However, if any participant encounters a problem, it rolls back the transaction and informs the coordinator who broadcasts an abort message to all participants.\\


        

    
    \subsection{Security module}
    This module encompasses a set of measures taken by SGX-based KVSs to provide security properties to their data even outside the enclave. It includes the CIA triad (\ie ``data confidentiality'', ``data integrity (with freshness)'', ``data availability'') and the enclave-inherent security property: enclave attestation and secure channel establishment.
    We explain below how SGX-based KVSs use enclaves to enforce each of the properties above even when data are not kept inside the enclave.

        \subsubsection{Enclave attestation and secure channel establishment} 
            \noindent\textbf{Enclave attestation.}
            Enclave attestation is used to verify the identity of a remote enclave (MRENCLAVE or MRSIGNER) by a client against an expected identity. This verification process assures the client that the remote enclave executes the expected code and can be relied upon for secure computation and data protection, establishing a foundation of trust in an untrusted environment. Enclave attestation is equivalent to certificate verification in PKI, where Intel acts as the main certification authority, with the additional verification of code integrity. As explained in section~\ref{ssec:remote-attestation}, there are two models for SGX RA: the DCAP model and the EPID model. All the surveyed SGX-based KVSs rely on the EPID model, even though it is deprecated. We believe this choice is due to its ease of use, as it only relies on IAS (Intel Attestation Service) without the need to deploy a dedicated infrastructure as required by the DCAP model to locally cache Intel certificates.\\
            \noindent\textbf{Secure channel establishment.}
            After attesting an enclave, the client and SGX server establish a secure channel, with the server-side endpoint inside the enclave. This ensures that a client can push (or pull) its data to (or from) the server enclave without disclosing it to the surrounding non-enclave environment. Intel's SGXSSL~\cite{web:sgxssl} or previous academic work such as TaLoS~\cite{talos} provide primitives to combine remote attestation (RA) and TLS to securely communicate with SGX servers.
            \\
    
        \subsubsection{Data confidentiality} 
        Data confidentiality includes protecting sensitive data from disclosure. In SGX-based KVSs, inherent confidentiality offered by enclave is combined with confidentiality offered by software encryption, since the enclave alone may not be large enough to keep all confidential data. Specifically, each SGX-based KVS develop its strategy to split its data structure across enclave and non-enclave environments, while using encryption key to cipher sensitive parts that are kept outside enclave, as explained in section~\ref{ssec:data-strcuture}. Since the encryption key does not leave the enclave, enclave confidentiality guarantees indirectly spans to non-enclave environments.
           
        \subsubsection{Data integrity and freshness} \label{sssec:integrity-freshness}
        Data integrity is usually associated with freshness to ensure that data is accurate, not tampered with and in its last version. Regarding in-enclave data, their integrity and freshness is ensured by design with the SGX built-in MHT for EPC pages (see section~\ref{sssec:memory-layout}). Regarding data that are kept outside the enclave, different methods are employed by SGX-based KVSs to guarantee their integrity and freshness:\\
        \noindent\textbf{Merkle Hash Tree.}
        MHT is a cryptographic structure used to ensure the integrity of large datasets, including KVSs. It constructs a tree-like structure over a KVS where each leaf node represents the hash of a KV pair, and each non-leaf node represents the hash of its child nodes' concatenated hashes. The root of the tree, known as the Merkle root, uniquely summarizes the entire KVS. ShieldStore~\cite{shieldstore} is an SGX-based KVS that utilizes MHT to ensure the integrity of its KV pairs. Due to the typically large size of MHTs, only the Merkle root is kept inside the enclave, serving as a ground truth reference for the state of the KV pairs. Speicher~\cite{speicher} also leverages one MHT per SSTables of same level to ensure their integrity and freshness.\\
        \noindent\textbf{SGX counters.}
        Built-in SGX counters are hardware-based monotonic counters provided by SGX technology to span the freshness protection provided by enclave to data located outside enclave boundaries, such as persisted data. These counters prevent replay and rollback attacks by offering a secure, tamper-resistant method for tracking the sequence of operations. By ensuring that each counter value associated with an operation or file is unique and incrementing, SGX counters help maintain data freshness. Furthermore, to detect data tampering, SGX counters are often paired with MAC computation, using an in-enclave MAC key. However, SGX counters are notoriously slow (60-250 ms)~\cite{rote} because they are synchronous and wear out after only a few weeks~\cite{speicher}. More importantly, Intel stopped supporting SGX monotonic counters since version 2.8 of the SDK in January 2020, leading developers to propose their own methods for building trusted counters. ShieldStore~\cite{shieldstore} is an SGX-based KVS that leveraged SGX counters to ensure the integrity and freshness of its persisted (sealed) data.\\
        \noindent\textbf{Custom software counters.}
        Since SGX monotonic counters are slow and have been deprecated by Intel, recent SGX-based KVSs have developed custom solutions for building trusted counters. For example, Speicher~\cite{speicher} and Treaty~\cite{treaty} systems designed an asynchronous trusted monotonic counter (AMC) to ensure the freshness and integrity of their log files. Each time data is modified, the counter is incremented asynchronously. This incrementation is only stabilized (persisted to a file) when the data itself is persisted. By separating the counter incrementation from its stabilization, AMC can maintain availability guarantees while optimizing performance, the incrementation being 7000x faster than the synchronous SGX counter. \\
        \noindent\textbf{Hash chain.}
        Hash chain aims to create a cryptographic link between successive entries of a log file to detect integrity violation. For instance, Tweezer~\cite{tweezer} leverages hash chain to maintain the integrity of its logs. Specifically, for each new log entry $e_i$, the system computes a MAC $M_i$ by concatenating the encrypted data entry with the previous log entry $(M_i = MAC( M_{i-1} || Enc(e_i) ))$, with $M_0$ being a random nonce. Freshness is ensured by using a new MAC key for each log entry, invalidating a replayed log that would be linked to an older MAC key. Tiks~\cite{engraft-tiks}, a distributed KVS, also utilizes hash chains for its logs and additionally leverages the broadcast protocol offered by Raft to maintain the consistency and freshness of the logs across multiple nodes.\\
        \noindent\textbf{Operation ID.} 
        Operation ID (oid) is an incremental nonce attached to messages over the network to guarantee their freshness. For instance, Precursor~\cite{precursor} keeps track, within its enclave, of each oid provided by a client for its KV packets to ensure that the packets are not replayed by an attacker. Avocado~\cite{avocado} and Treaty~\cite{treaty}, both distributed KVSs, also leverage authenticated packets with oid to track requests/responses and transactions, respectively.\\

        \subsubsection{Data availability} 
        Data availability ensures that the data is available when needed, despite the presence of failures (\eg crashes), including enclave failures. To ensure this property, SGX-based KVSs rely on three methods :\\
        \noindent\textbf{Persistency.} 
        Persistency refers to the act of saving data from volatile memory (RAM) to non-volatile storage (disk), such as flushing in-memory MemTable to disk in form of SSTables, in the context of LSMT data structure. 
        Persistency is natively used by persistent SGX-based KVSs such as Speicher~\cite{speicher}, Tweezer~\cite{tweezer} and Treaty~\cite{treaty}. However, to keep in-enclave data, such as software encryption keys, secure (in terms of confidentiality and integrity) even after enclave is shutdown, they rely on SGX sealing.  \\
        \noindent\textbf{Logging.} 
        Logging involves recording every write operation in a KVS sequentially in log files before the operation is applied to the main data structure. The goal is to ensure immediate durability by enabling data replay in case a failure occurs before the data is pushed to the main data structure. Such mechanism is generally employed by log-based KVSs (\ie those that rely on LSMT~\cite{speicher,tweezer,treaty}). Logging is usually combined with persistency to allow logs to be refreshed after persistency, by eliminating entries of persisted data from logs.\\
        \noindent\textbf{Replication.} 
        Replication is a mechanism usually leveraged by distributed KVSs that do not support data persistency (\eg Avocado~\cite{avocado}) to store copies of data on other machines. To ensure the consistency and integrity of replication, consensus mechanisms (see section~\ref{ssec:synchronization-module}) are employed across nodes. The main advantage of replication over persistency as a data availability strategy is that it eliminates the additional latency induced by flushing data to disk.

\section{Classification of TEE-based KVSs} \label{sec:classification}
    As shown in Table~\ref{tab:surveyed-kvs}, we categorize the surveyed KVS into single-node KVS or multi-node (distributed) KVS. Single-node KVSs are databases that store KV pairs on a single machine while multi-node KVS store KV pairs on multiple machines through replication, sharding or both. We recall that all the surveyed SGX-based KVSs made the choice to use client SGX, which has a limited secure memory of 128MB, as their TEE implementation. We classify surveyed KVSs according to to three criteria: (1) Data access, \ie \textit{the underlying data structure used while outlining KVSs that support range queries}; (2) security, \ie \textit{how data are partitioned between enclave/non-enclave environments, what strategies are deployed to ensure data integrity with freshness outside the enclave, and what strategies are used to guarantee availability of data}; (3) communication, \ie \textit{what are the employed I/O optimization techniques and the consensus algorithms used in distributed context}.

\begin{table}
\centering
\renewcommand{\arraystretch}{1.5}  
\scriptsize
        \begin{tabularx}{0.48\textwidth}{|X|X|} 
        \hline
        \multicolumn{1}{|c|}{\textbf{Single-node KVSs}} & \multicolumn{1}{|c|}{\textbf{Multi-node KVSs}}\\
        \hline
         VeritasDB~\cite{veritasDB}, ShieldStore~\cite{shieldstore}, Speicher~\cite{speicher}, Precursor~\cite{precursor}, Tweezer~\cite{tweezer}, Concerto~\cite{concerto}, eLSM~\cite{elsm} & Avocado~\cite{avocado}, Treaty~\cite{treaty}, Tiks~\cite{engraft-tiks} \\
        \hline
        \end{tabularx}
\captionsetup{font=small} 
\caption{Surveyed TEE-based KVSs}
\label{tab:surveyed-kvs}    
\end{table}

    \subsection{Data access criteria}
        We have previously enumerated the data structures commonly used in SGX-based KVSs (see section~\ref{ssec:data-strcuture}). In this section, we classify existing SGX-based KVSs in Table~\ref{tab:data-access} according to their underlying data structure. Frameworks that act as proxy atop existing KVS such as VeritasDB~\cite{veritasDB} and Concerto~\cite{concerto} can use any data structure to store KV pairs with no or minimal changes, so they are not classified. Table~\ref{tab:data-access} also outlines KVSs that support range query operations in addition to the classical \textit{get(K)/put(K, V)}.
        Below, we explain how each SGX-based KVS customized its data structure to exploit SGX features while addressing the secure memory size limitation.\\
        \noindent\textbf{Hash Table.}
        ShieldStore~\cite{shieldstore} leverages a hash table to store its KV pairs. To avoid stressing the limited EPC, the hash table is kept outside the enclave with encrypted KV pairs. Only the bare minimum data is kept inside the enclave, such as cryptographic keys and MHT root.
        Precursor also leverages uses a hash table put prefer to keep the keys part in enclave and only the encrypted values part outside enclave.\\
        \noindent\textbf{Skip list.}
        Avocado~\cite{avocado} is an in-memory distributed SGX-based KVS that employs skip-list on each individual node to store its KV pairs. The skip list spans across enclave and non-enclave environment. The skip list index, keys, pointers to values and their individual hashes are kept inside the enclave. The values are stored encrypted outside the enclave to alleviate the impact on the EPC.\\
        \noindent\textbf{LSMT.} 
        Speicher~\cite{speicher} is a pioneering persistent SGX-based KVS that leverages LSMT as its underlying data structure. It uses a skip list for its MemTable implementation, partitioned between the enclave and non-enclave environment in the same manner as Avocado~\cite{avocado}. Each SSTable is stored on disk (\ie outside the enclave) and authenticated using one per-block hash and a global hash over the per-block hashes. The entire SSTables are authenticated using a MHT whose root is kept inside the enclave.
        Tweezer~\cite{tweezer} builds upon Speicher and proposes two design optimizations: using a unique MAC key per SSTable instead of building an MHT over SSTables to improve latency, and using finer-grained authenticated SSTables with one MAC per KV pair to save EPC memory during read operations instead of one hash per SSTable block.
        Treaty~\cite{treaty} also leverages Speicher as a building block to construct a distributed transactional SGX-based KVS by adding a distributed transaction layer (2PC) to support ACID transaction over multiple replicated nodes.
        Another noteworthy work is eLSM~\cite{elsm}, which acts as an add-on (middleware) to existing non-SGX LSMT-based KVSs such as RocksDB~\cite{web:rocksdb} and LevelDB~\cite{web:leveldb}. eLSM provides an enclave environment to store the logic of the KVS, index structure, and authentication data such as the MHT root of SSTables, with minimal changes to off-the-shelf LSMT-based KVSs.\\

\begin{table}[!h]
\centering
\scriptsize
\renewcommand{\arraystretch}{1.5}  
    \begin{tabularx}{0.48\textwidth}{|X|X|X|} 
    \hline
    \multicolumn{3}{|c|}{\textbf{Data structures}} \\
    \hline
    \multicolumn{1}{|c|}{\textbf{Hash table}} & \multicolumn{1}{|c|}{\textbf{LSMT}} & \multicolumn{1}{|c|}{\textbf{Skip list}}\\
    \hline
     \multicolumn{1}{|>{\centering\arraybackslash}X|}{\cite{shieldstore,precursor}} & \multicolumn{1}{|>{\centering\arraybackslash}X|}{\cite{speicher,treaty,tweezer,elsm}} & \multicolumn{1}{|>{\centering\arraybackslash}X|}{\cite{avocado}} \\
    \hline
    \end{tabularx}

    \begin{tabularx}{0.48\textwidth}{|X|} 
    \hline
    \multicolumn{1}{|c|}{\textbf{Range queries support}} \\
    \hline
    \multicolumn{1}{|>{\centering\arraybackslash}X|}{\cite{speicher,tweezer,elsm}} \\
    \hline
    \end{tabularx}
\captionsetup{font=small} 
\caption{Data Access}
\label{tab:data-access}
\end{table}

            
    \subsection{Security criteria}
    In this criterion, we classify SGX-based KVSs according to the security measures used to ensure the confidentiality, integrity (with freshness), and availability of data, by referring to Table~\ref{tab:security}. The table is subdivided into three parts:\\
    \noindent\textbf{Data separation.} 
    This part highlights the generic partitioning of data between the enclave environment and the non-enclave environment, adopted by all SGX-based KVSs to overcome the SGX secure memory limitation. Data kept outside the enclave are generally large buffers that would consume significant secure memory. Being outside the enclave, they necessitate additional measures to secure them. Data kept inside the enclave are inherently secure regarding confidentiality and integrity (with freshness). This includes, among other elements, authentication data (hashes, MACs, etc.) and encryption keys to authenticate and encrypt/decrypt data located outside the enclave.
    \\
    \noindent\textbf{Integrity and freshness of outside-enclave data.}
    This part focuses on the different methods used to provide integrity and freshness of data kept outside the enclave. We do not discuss ensuring the confidentiality of these data, as it is systematically achieved using encryption algorithms such as AES-GCM, with the encryption key kept inside the enclave to prevent unauthorized decryption.
    \\
    \noindent\textbf{Availability mechanisms.}
    This part focuses on the different methods used by SGX-based KVSs to ensure that their KV pairs remain available even in the event of crash failures.
    \\

\begin{table*}
\centering
\renewcommand{\arraystretch}{1.5}  
\scriptsize
    \begin{tabularx}{\textwidth}{|p{0.5\textwidth}|X|} 
    \hline
    \multicolumn{2}{|c|}{\textbf{Data separation}} \\
    \hline
    \multicolumn{1}{|c|}{\textbf{In-enclave data}} & \multicolumn{1}{|c|}{\textbf{Outside-enclave data}} \\
    \hline
    \parbox{0.5\textwidth}{
        \vspace{3pt}
        KVS operations logic,\\ 
        I/O optimization logic (SPDK,DPDK libraries),\\ 
        Consensus logic,\\ 
        Index structure (excluding~\cite{veritasDB,concerto}), \\ 
        Pointers to external data (\eg in-memory values),\\ 
        Cryptographic keys (for encryption, MAC, session),\\ 
        Hashes of in-memory data (\eg values) or MHT root,\\ 
        Trusted counters,\\ 
        Cached KV pairs
        \vspace{3pt}} 
    & 
    \parbox{0.5\textwidth}{
      Encrypted Vs (or KV pairs~\cite{veritasDB,shieldstore,concerto}),\\ 
      MHT,\\ 
      Log files and persisted data (\eg SSTables),\\ 
      SPDK/DPDK buffers} \\
    \hline
    \end{tabularx}

    \begin{tabularx}{\textwidth}{|X|X|X|X|X|} 
    \hline
    \multicolumn{5}{|c|}{\textbf{Integrity and freshness of outside-enclave data}} \\
    \hline
    \multicolumn{1}{|c|}{\textbf{Merkle Tree}} & \multicolumn{1}{|c|}{\textbf{Hash chain}} & \multicolumn{1}{|c|}{\textbf{SGX counters}} & \multicolumn{1}{|c|}{\textbf{Custom software counters}} & \multicolumn{1}{|c|}{\textbf{Operation id}} \\
    \hline
    \multicolumn{1}{|>{\centering\arraybackslash}X|}{\cite{veritasDB,shieldstore,speicher,treaty}} & \multicolumn{1}{|>{\centering\arraybackslash}X|}{\cite{tweezer,engraft-tiks}} & \multicolumn{1}{|>{\centering\arraybackslash}X|}{\cite{veritasDB,shieldstore}} & \multicolumn{1}{|>{\centering\arraybackslash}X|}{\cite{speicher,treaty}} & \multicolumn{1}{|>{\centering\arraybackslash}X|}{\cite{precursor,avocado,treaty}} \\
    \hline
    \end{tabularx}

    \begin{tabularx}{\textwidth}{|X|X|X|} 
    \hline
    \multicolumn{3}{|c|}{\textbf{Availability mechanisms}} \\
    \hline
    \multicolumn{1}{|c|}{\textbf{Persistency}} & \multicolumn{1}{|c|}{\textbf{Logging}} & \multicolumn{1}{|c|}{\textbf{Replication}} \\
    \hline
    \multicolumn{1}{|>{\centering\arraybackslash}X|}{\cite{veritasDB,shieldstore,speicher,treaty,elsm,tweezer,engraft-tiks}} & \multicolumn{1}{|>{\centering\arraybackslash}X|}{\cite{speicher,treaty,elsm,tweezer,engraft-tiks}} & \multicolumn{1}{|>{\centering\arraybackslash}X|}{\cite{avocado,engraft-tiks,treaty}} \\
    \hline
    \end{tabularx}
\captionsetup{font=small} 
\caption{Security methods}
\label{tab:security}
\end{table*}

    \subsubsection{Data separation}
    Regardless of the studied SGX-based KVS, a generic framework is followed to partition data between enclave and non-enclave environments, as shown in Table~\ref{tab:security}. Specifically, we do not focus on what elements each specific KVS has, but rather provide a comprehensive overview showing the location of each element, if present, in an SGX-based KVS. \\
    \noindent\textbf{In-enclave data.}
    Regarding in-enclave data, we find the different application's logic to ensure their execution reliability: 
    (1) KVS operations logic (\eg \textit{get/put}, compaction, \etc);
    (2) I/O optimization logic (SPDK, DPDK libraries) and;
    (3) Consensus logic employed in distributed KVSs.
    Index structures (including keys of KV pairs) are generally also stored inside the enclave to ensure reliable data fetching. Some exceptions, like VeritasDB~\cite{veritasDB} and Concerto~\cite{concerto}, keep their index structure outside the enclave to save even more secure memory. If values are kept outside the enclave, their pointers, alongside their hashes, are maintained inside the enclave to detect any tampering with the values. The root of big authentication data (\eg MHT) is also kept inside the enclave as a reference for trust. Cryptographic keys used to encrypt, compute MACs for outside-enclave data, or secure sessions between clients and enclaves, are stored within the enclave to prevent unauthorized data decryption, alteration, or session spoofing. Finally, some frameworks, like VeritasDB~\cite{veritasDB}, may also dedicate a secure memory portion to cache their latest accessed KV pairs to avoid re-authenticating them in case they are accessed again. 
    \\
    \noindent\textbf{Outside-enclave data.}
    Regarding outside-enclave data, we find the encrypted values (or the encrypted KV pairs) as their size may be prohibitive for the enclave. Large authentication data structures, such as the MHT (excluding the root), are also kept outside the enclave. Additionally, large data buffers, such as those used in SPDK and DPDK, are stored outside the enclave for the same reason. Finally, on-disk files, including persisted data and logs, are inherently located outside the enclave.\\

    \subsubsection{Integrity and freshness of outside-enclave data}
    \noindent\textbf{Merkle tree.}
    Methods that leverage Merkle Hash Trees (MHT) to safeguard the integrity and freshness of their data keep the MHT root inside the enclave. For instance, VeritasDB~\cite{veritasDB} and ShieldStore~\cite{shieldstore} use MHT to secure the KV pairs of their underlying data structures, while Speicher~\cite{speicher} and Treaty~\cite{treaty} use one MHT per SSTable level to secure their SSTables.\\
    \noindent\textbf{Hash chain.}
    Tweezer~\cite{tweezer} and Tiks~\cite{engraft-tiks} both leverage hash chain to secure the integrity of their logs data. For ensuring their freshness, Tweezer utilizes a fresh in-enclave MAC key to authenticate each new log entry, while Tiks rely on its distributed consensus (Raft) for that (see section~\ref{ssec:synchronization-module}). Details are provided in \textit{\ref{sssec:integrity-freshness}: Hash chain}.\\
    \noindent\textbf{SGX counters.}
    VeritasDB~\cite{veritasDB} and ShieldStore~\cite{shieldstore} both leverage SGX built-in counters to ensure the freshness of their sealed enclave data. However, as mentioned in \textit{\ref{sssec:integrity-freshness}: SGX counters}, Intel has deprecated the use of SGX counters. Consequently, these methods need to change their approach to ensure freshness.\\
    \noindent\textbf{Custom software counters.}
    Speicher~\cite{speicher} and its distributed version, Treaty~\cite{treaty}, leverage custom in-enclave asynchronous monotonic counters to ensure freshness of their log files. Details are given in \textit{\ref{sssec:integrity-freshness}: Custom software counters}.\\
    \noindent\textbf{Operation id.}
    Precursor~\cite{precursor}, Avocado~\cite{avocado} and Treaty~\cite{treaty} are SGX-based KVSs that explicitly mention tracking packet versions received through the network, inside enclave, to detect replayed packets. 

    \subsubsection{Availability mechanisms}
    All LSMT-based KVSs~\cite{speicher,treaty,tweezer,elsm} leverage persistency and logging by design to write their SSTables and keep track of not-yet-persisted KV pairs. Other SGX-based KVSs~\cite{veritasDB,shieldstore} implement their own mechanisms of persistency atop their storage. Finally, distributed KVSs~\cite{avocado,treaty,engraft-tiks} rely either on lone replication~\cite{avocado,engraft-tiks} to keep their data available or on all the previous methods~\cite{treaty} to improve availability guarantees.

    \subsection{Communication criteria}
    Table~\ref{tab:communication} classifies SGX-based KVSs according to the mechanisms adopted to communicate with peripherals (storage disks, network cards) or other KVS nodes (to establish a consensus) from within the enclave. 

\begin{table}[!h]
\centering
\renewcommand{\arraystretch}{1.5}  
\scriptsize
    \begin{tabularx}{0.48\textwidth}{|X|X|} 
    \hline
    \multicolumn{2}{|c|}{\textbf{Disk access optimization}} \\
    \hline
    \multicolumn{1}{|c|}{\textbf{Async syscall}} & \multicolumn{1}{|c|}{\textbf{SPDK+DMA}} \\
    \hline
    \multicolumn{1}{|>{\centering\arraybackslash}X|}{\cite{treaty,engraft-tiks}} & \multicolumn{1}{|>{\centering\arraybackslash}X|}{\cite{speicher}} \\
    \hline
    \end{tabularx}

    \begin{tabularx}{0.48\textwidth}{|X|X|} 
    \hline
    \multicolumn{2}{|c|}{\textbf{Networking optimization}} \\
    \hline
    \multicolumn{1}{|c|}{\textbf{Async syscall}} & \multicolumn{1}{|c|}{\textbf{DPDK+RDMA}} \\
    \hline
    \multicolumn{1}{|>{\centering\arraybackslash}X|}{\cite{engraft-tiks}} & \multicolumn{1}{|>{\centering\arraybackslash}X|}{\cite{avocado,treaty,precursor}} \\
    
    \hline
    \end{tabularx}

    \begin{tabularx}{0.48\textwidth}{|X|X|X|} 
    \hline
    \multicolumn{3}{|c|}{\textbf{Synchronization}} \\
    \hline
    \multicolumn{1}{|c|}{\textbf{Raft}} & \multicolumn{1}{|c|}{\textbf{ABD}} & \multicolumn{1}{|c|}{\textbf{2PC}} \\
    \hline
    \multicolumn{1}{|>{\centering\arraybackslash}X|}{\cite{engraft-tiks}} & \multicolumn{1}{|>{\centering\arraybackslash}X|}{\cite{avocado}} & \multicolumn{1}{|>{\centering\arraybackslash}X|}{\cite{treaty}} \\
    \hline
    \end{tabularx}
\captionsetup{font=small} 
\caption{Communication methods}
\label{tab:communication}
\end{table}

\section{Side-channel leakage} \label{sec:sca}
Despite SGX's security properties, it is far from perfect, as its popularity makes it a target for various attacks, especially side-channel attacks (SCAs)~\cite{sca}. The latter represent a class of security threats that exploit unintended information leakage from a system's physical implementation rather than its theoretical cryptographic weaknesses. SCAs leverage various vectors of information leakage, among others: (1) page access or cache access pattern leakage by analyzing page accesses, page faults and cache hit/miss; (2) enclave interface invocation leakage by analyzing enclave function invocation delay; (3) instruction trace leakage by analyzing leaked instruction sizes; (4) volume leakage by analyzing the size (or volume) of the query result~\cite{volume-leakage,volume-leakage-2,volume-leakage-3,volume-leakage-4,volume-leakage-5,volume-leakage-6}.
By analyzing the previous leaked information, attackers can glean valuable information about the internal state of the system, including secret keys or other sensitive data. 

\subsection{Classification w.r.t compromised security objectives}
A previous survey~\cite{sgx-vulnerabilities} has classified SCAs into three categories based on the SGX security objectives that may be compromised: 
    \noindent\textbf{Confidentiality impairment:} SCAs that falls under this category can obtain secrets from ISV enclave (\ie application enclave) instances, where KV pairs may be stored~\cite{sgx-co-1,sgx-co-2,sgx-co-3,sgx-co-4,sgx-co-5,sgx-co-6,sgx-co-7,sgx-co-8,sgx-co-9,sgx-co-10,sgx-co-12,sgx-co-13,sgx-co-14,volume-leakage,volume-leakage-2,volume-leakage-3,volume-leakage-4}.\\
    \noindent\textbf{Attestation security impairment:} SCAs that falls under this category can obtain secrets from architectural enclave instances, \ie Quoting enclaves (see section~\ref{ssec:remote-attestation}), compromise the integrity of the attestation mechanism and thus impersonate enclaves~\cite{sgx-as-1,sgx-as-2,sgx-as-3,sgx-as-4}. \\
    \noindent\textbf{Usability impairment:} SCAs that falls under this category can interrupt the work of enclaves through Denial-of-Service (DoS) and render them unresponsive~\cite{sgx-bomb}.
It is worth noting that some SCAs may fall under multiple categories, usually causing both confidentiality and attestation security impairments simultaneously. Examples include SGXPectre~\cite{sgx-as-2}, an SGX-variant of the Spectre attack~\cite{spectre}, and Foreshadow~\cite{sgx-as-4}, a Meltdown-style attack~\cite{meltdown} against SGX.

\subsection{Mitigation}
Existing countermeasures to SCAs can be categorized into three classes: hardware solutions, application solutions and system solutions.

\subsubsection{Hardware solutions} 
Hardware solutions require modifications to processors and their microcode, which can only be implemented by manufacturers. For instance, architectural changes introduced by Cascade Lake CPUs enabled Intel to provide effective fixes against both SGXPectre and Foreshadow~\cite{spectre-meltdown-fix}. However, these solutions require a long time window before deployment in commercial CPUs, and only CPU manufacturers (\eg Intel, AMD, ARM) can propose these fixes. 

\subsubsection{Application solutions}
Application solutions are proposed by literature to mitigate SCAs before CPU manufacturers propose hardware mitigation. They are usually hardware-agnostic and standalone security layers that may interface with existing applications. We leverage~\cite{sgx-vulnerabilities} and recent literature about volume leakage attacks to classify application solutions into five categories according to their defense strategy and mitigated attacks.

\noindent\textbf{Deterministic multiplexing.}
Deterministic multiplexing (DM)~\cite{sgx-co-1} protects SGX-enabled systems from page-fault driven attacks. It leverages page-fault obliviousness, ensuring input data does not affect the number of allocated pages, thus preventing sensitive information leakage through page faults. However, the software implementation of DM could lead to an average overhead of 705x, while a dedicated hardware implementation reduce overhead to 6.77\%.\\
\noindent\textbf{Anomaly detection.}
Similar to DM, these methods aim to mitigate page-fault driven SCAs. They are based on page anomaly detection as an indicator of on-going SCA. They rely on Intel’s hardware transactional memory (TSX)~\cite{web:tsx} to handle page exceptions and interrupts, without the involvement of the OS. 
Specifically, T-SGX~\cite{t-sgx} redirects exceptions to a specific page, allowing a user-space fallback handler to manage errors, bypassing the underlying OS. T-SGX increased execution time by 40\% and memory usage by 30\%.
Déjà vu~\cite{deja-vu} used TSX to create a reliable time measurement tool for anomaly detection, independent from the OS, effectively countering page-fault attacks despite increasing runtime.
Regarding cache-access driven SCA, Cloak~\cite{cloak} used TSX to ensure that critical data stays in the CPU cache during transactions, proving effective with minimal overhead (1.2\%) for low-memory tasks but significant overhead (248\%) for memory-intensive ones.\\
\noindent\textbf{Randomization.}
randomization is a memory protection technique that mitigates page-fault driven SCAs attacks by randomly loading data and code in the OS, making it difficult for attackers to locate targets. 
SGX-Shield~\cite{sgx-shield}, a code randomization technique, uses a secure in-enclave loader to secretly randomize memory layout with fine granularity by splitting target code into small randomization units and loading each at random addresses.
DR.SGX~\cite{drsgx}, a data location randomization technique, breaks the link between an attacker’s memory observations and a victim’s data access patterns. It permutes data locations at a fine granularity using small-domain encryption and the CPU’s AES-NI hardware acceleration. To prevent correlation attacks from repetitive access patterns, periodic re-randomization of enclave data is applied, although this results in performance overheads ranging from 4.36× to 11× depending on the re-randomization rate.\\
\noindent\textbf{Volume leakage mitigation.}
In the context of KVSs, an attacker can monitor value size for simple \textit{get} operations and/or the number of matched keys for \textit{get-range} operations~\cite{volume-leakage-2,volume-leakage-3}. For multi-maps, where each key is associated with a vector of values, the adversary can also analyze the number of values associated with a specific key~\cite{volume-leakage-5,volume-leakage-6}. Discussed methods to mitigate these attacks~\cite{volume-leakage-3,volume-leakage-4} include: (1) padding or injecting random noise (differential privacy) to query results or database to conceal the actual number of records or their size; (2) preventing query replay by uniquely identifying and recording each request; (3) preventing data injection by filtering suspicious data during \textit{put} operations. However, these methods can introduce significant server overhead and impact user experience. 
Recent works aim to minimize server storage overhead caused by naive padding while still hiding volume information. For instance, dprfMM~\cite{volume-leakage-5} uses a cuckoo hash-based method, and XorMM~\cite{volume-leakage-6} employs an XOR filter to support volume hiding in multi-maps while minimizing server storage overhead. Similarly, Veil~\cite{veil} randomly assigns each key to a set of equally sized buckets, which are padded if needed and may overlap for different keys. 
\\
\noindent\textbf{ORAM.}
ORAM (Oblivious RAM)~\cite{oram} is designed to protect from nearly all SCA vectors. It is a cryptographic construct that ensures secure access to memory regions on untrusted servers by accessing multiple locations per operation and re-shuffling/re-encrypting memory with a random seed. 
OBFUSCURO~\cite{obfuscuro} uses SGX for program obfuscation by enforcing code execution and data access via ORAM operations. It transforms the program layout to be ORAM-compatible and ensures fixed time intervals for program runs. Still, OBFUSCRO incurs significant run-time performance overhead (83× on average), although it is faster than most cryptographic obfuscation schemes.\\

\subsubsection{System solutions}
System solutions modify existing system software by integrating tailored SCA mitigation in their design. 
A notable example in databases is Opaque~\cite{opaque}, an oblivious SGX-based distributed data analytics platform built on top of Spark SQL. At a basic level, Opaque leverages SGX to provide hardware-based data encryption, attestation, and enforce correct execution. In its oblivious mode, Opaque additionally provides oblivious execution, which protects against access pattern leakage in SQL queries. Opaque's method differs from Oblivious RAM (ORAM) as it takes into account specific SQL operations such as sorting. In its oblivious pad mode, Opaque pads the final output to prevent size leakage based SCAs. However, obliviousness is fundamentally costly, and Opaque's oblivious execution incurs an overhead of up to 46x compared to non-oblivious execution.
ObliDB~\cite{oblidb} further provides obliviousness while being 19x faster than Opaque, but still does not reach practical performance.

\section{Discussion} \label{sec:discussion} 

    \noindent\textbf{Preference of Client SGX over Scalable SGX.}
    All the TEE-based KVSs we surveyed use SGX implementations, specifically the client version of SGX, which offers the most extensive set of security features, including resilience against physical integrity attacks. However, this version has a small secure memory (128MB), requiring developers to adopt novel designs to avoid costly page swapping.
    Scalable SGX overcomes the limited secure memory size by providing an EPC size up to 512GB, reducing the effort needed to port applications to a secure environment. However, Intel achieved this by removing the built-in MHT, which protects against physical integrity attacks in the EPC. Without this protection, an attacker can mount replay attacks, replacing memory content with previous versions undetected. For instance, an attacker could revert a security hot-patch to reintroduce a vulnerability, allowing them to extract enclave secrets.
    To reconcile the large EPC size of Scalable SGX with the safety of Client SGX, Aublin et al.~\cite{sgx-integrity} propose a PoC that leverages both SGX versions: Scalable SGX for hosting and executing the main application code and Client SGX for hosting verification metadata of Scalable SGX pages.
    However, it is not clear whether Intel will keep supporting Client SGX alongside Scalable SGX. Otherwise, integrity protection of Scalable SGX against physcial attacks should be delegated to dedicated hardware such as on-chip hardware~\cite{keystone} or FPGA~\cite{penglai}. 
    \\
    
    \noindent\textbf{Continuous problem of SCAs.}
    Intel prioritizes patching SCAs that threaten the attestation mechanism of SGX, as these attacks undermine the core security features of SGX. For example, Intel mitigated Foreshadow~\cite{sgx-as-4} and the most dangerous Spectre variants~\cite{sgx-as-2} through microcode updates~\cite{spectre-meltdown-fix}. However, for other SCAs that may leak end-user secrets without compromising SGX's built-in attestation secrets, Intel asserts that \textit{``it is the enclave developer's responsibility to address side-channel attack concerns''}~\cite{web:sgx-developer-guide}. 
    In the absence of Intel's mitigation, application-level solutions provide generic defenses against various SCAs. However, these solutions are expensive to implement, and thus none of the surveyed KVSs employed them, as they prioritize performance for real-world viability. Additionally, SCAs leverage unexpected channels, complicating future prevention efforts. The Keccak Team notes, \textit{``protection against side-channel attacks is never expected to be absolute: a determined attacker with massive resources will sooner or later break an implementation. The engineering challenge is to implement enough countermeasures to make the attack too expensive to be interesting''}~\cite{web:keccak-team}. Thus, future research will likely continue to focus on mitigating existing SCAs while controlling performance loss. An emerging approach may involve using specifically designed enclave oblivious memory~\cite{sanctum,ghost-rider} for building KVSs.
    \\

\section{Conclusion} \label{sec:conclusion}
    SGX-enabled CPUs provide security and trust assurances for various computing systems and services, including storage services like KVSs that handle sensitive data. However, integrating SGX introduces performance challenges due to limited secure memory and context switching, complicating the design of SGX-based KVSs. This survey aims to elucidate the complexities of SGX-based KVS design by identifying various modules that collectively form a fully functional KVS that attempt to address both security and performance considerations. We classify existing SGX-based KVSs according to multiple criteria reflecting these modules, providing an overview of the diverse strategies adopted in the literature to build secure and practical KVSs. Additionally, we highlight the persistent threat of SCAs to SGX systems, classify these attacks, and discuss available mitigations. Although Intel continuously patches the most dangerous SCAs affecting SGX attestation, mitigations for SCAs impacting end-user applications are left to enclave developers. Proposed mitigations effectively address access pattern leakage, a common SCA vector, but often result in impractical performance. Thus, none of the surveyed SGX-based KVSs implemented SCA mitigations, leaving them vulnerable to attacks that could leak users' KV pairs even with SGX protection. Future SGX-based KVSs will need to address this issue while preserving practical performance, suggesting the need for novel enclave architectures. Another concern is whether Intel will continue supporting Client SGX, the TEE implementation with the most comprehensive security features, or shift focus to Scalable SGX to meet the growing server demand for larger and faster secure memory, albeit at the expense of a broader security spectrum.

\section*{Acknowledgments}
This work was supported by a French government grant managed by the Agence Nationale de la Recherche under the France 2030 program, reference “ANR-23-PECL-0007" as well as the ANR Labcom program, reference "ANR-21-LCV1-0012".

\bibliographystyle{spmpsci}
\bibliography{bibliography}

\end{document}